\documentclass[journal]{IEEEtran}
\usepackage{amsmath,amsfonts}
\usepackage{array}
\usepackage{booktabs}
\usepackage{algpseudocode}
\usepackage{algorithm}
\usepackage{textcomp}
\usepackage{stfloats}
\usepackage{url}
\usepackage{verbatim}
\usepackage{graphicx}
\usepackage[font=footnotesize]{caption}
\usepackage{subcaption}
\usepackage{supertabular}
\usepackage{multirow}
\usepackage{placeins}
\usepackage{makecell}
\usepackage{adjustbox} 
\usepackage{algorithm,algpseudocode}
\usepackage[backend=biber, sorting=none]{biblatex}
\usepackage{amsmath}

\newtheorem{remark}{\bf Remark}

\newcommand{\Ccap}[1]{\text{C}_{\text{cap}}}

\newcommand{\RSD}[1]{\text{R}_{\text{SD}}}
\newcommand{\RS}[1]{\text{R}_{\text{S}}}
\newcommand{\RSH}[1]{\text{R}_{\text{SH}}}
\newcommand{\RSE}[1]{\text{R}_{\text{SE}}}

\newcommand{\Vsoc}[1]{{V}_{\text{SoC}}}
\newcommand{\VOC}[1]{{V}_{\text{OC}}}
\newcommand{\VSH}[1]{V_\text{SH}}
\newcommand{\VPV}[1]{V_\text{PV}}
\newcommand{\VD}[1]{V_\text{D}}
\newcommand{\VBatt}[1]{V_\text{Bt}}

\newcommand{\IBatt}[1]{{I}_{\text{Bt}}}
\newcommand{\ID}[1]{{I}_{\text{D}}}
\newcommand{\IPH}[1]{{I}_{\text{PH}}}
\newcommand{\Izero}[1]{{I}_{\text{0}}}
\newcommand{\ISH}[1]{{I}_{\text{SH}}}
\newcommand{\IS}[1]{{I}_{\text{S}}}
\newcommand{\IPV}[1]{{I}_{\text{PV}}}
\newcommand{\ItildG}[1]{\Tilde{I}^{\text{G}}_k}
\newcommand{\ItildL}[1]{\Tilde{I}^{\text{L}}_k}
\newcommand{\ItildPV}[1]{\Tilde{I}_{\text{PV}}}
\newcommand{\ItildBatt}[1]{\Tilde{I}_{\text{Bt}}}

\newcommand{\Zload}[1]{${Z}^{load}$}
\newcommand{\zpvi}[1]{{I}_\text{PV,z}}
\newcommand{\zpvv}[1]{{V}_\text{PV,z}}
\newcommand{\zphi}[1]{{I}_\text{PH,z}}
\newcommand{\zbatti}[1]{{I}_\text{Bt,z}}
\newcommand{\zbattv}[1]{{V}_\text{Bt,z}}

\newcommand{\npvv}[1]{n^\text{PV}_{V}}
\newcommand{\npvi}[1]{n^\text{PV}_{I}}
\newcommand{\nphi}[1]{n^\text{PH}_{I}}
\newcommand{\nbattv}[1]{n^\text{Bt}_{V}}
\newcommand{\nbatti}[1]{n^\text{Bt}_{I}}
\newcommand{\nrtur}[1]{n^\text{r}_{k}}
\newcommand{\nrtui}[1]{n^\text{i}_{k}}
\newcommand{\PGk}[1]{P^{\text{G}}_k}
\newcommand{\QGk}[1]{Q^{\text{G}}_k}
\newcommand{\PLk}[1]{P^{\text{L}}_k}
\newcommand{\QLk}[1]{Q^{\text{L}}_k}
\newcommand{\Vmagk}[1]{|V|^{\text{G}}_k}

\newcommand{\Gkz}[1]{G_{z,k}}
\newcommand{\Bkz}[1]{B_{z,k}}

\newcommand{\cktse}[1]{\textbf{ckt-SE}}
\newcommand{\pvse}[1]{\textbf{PV-SE}}
\newcommand{\btse}[1]{\textbf{Bt-SE}}

\newcommand{\cktpse}[1]{\textbf{ckt-PSE}}

\newcommand{\cktser}[1]{\textbf{ckt-SE}^\textbf{Re}}
\newcommand{\cktpser}[1]{\textbf{ckt-PSE}^\textbf{Re}}

\newcommand{\case}[2]{%
  \ifnum#1=1 TC1%
  \else\ifnum#1=2 TC2#2%
  \else\ifnum#1=3 TC3%
  \fi\fi\fi
}
\def\thanksto#1{
\begingroup
\def\thefootnote{}
\footnote{
\kern -3pt
\hrule width 0.4\columnwidth height 0.2pt
\kern 5pt
#1
}
\setcounter{footnote}{0}%\let\thanksto\relax\let\@thanksto\relax
\endgroup
}
\title{Circuit-Theoretic Joint Parameter-State Estimation of Utility-Scale Photovoltaic, Battery, and Grid Systems}

\author{
\begin{tabular}{c c}
 \textbf{Peng Sang} &  \textbf{Amritanshu Pandey} \\
 The University of Vermont &  The University of Vermont \\
 \underline{peng.sang@uvm.edu} &  \underline{amritanshu.pandey@uvm.edu} \\
\end{tabular}
}

\date{}

\begin{document}

\maketitle

\begin{abstract}
Solar PV and battery storage systems have become integral to modern power grids. 
Therefore, bulk grid models in real-time operation must include their physical behavior accurately for analysis and optimization.
AC state estimation is critical to building real-time bulk power systems models.
However, current ACSE techniques do not include detailed physics and measurements for battery and PV systems.
This results in sub-optimal estimation results and subsequent less accurate bulk grid models for real-time operation.

To address these challenges, we formulate a circuit-theoretic AC state estimator with accurate PV and battery systems physics and corresponding measurements.
First, we propose an aggregated equivalent circuit model of the solar PV, battery, and traditional grid components. 
Next, we add measurements from PV and battery systems to the traditional measurement set to facilitate accurate estimation of the overall grid model. 
Finally, we develop a circuit-theoretic joint parameter-state estimation algorithm that can accurately estimate grid, PV, and battery system states and is robust against erroneous parameters. 
To demonstrate the efficacy of the proposed framework, we estimate the states of 10k node transmission networks with hundreds of battery+PV-tied systems. 
We compare the accuracy against the estimation of stand-alone grid, battery, and PV systems.
%\TODO{Add one sentence to cover the experimental design, up to what size case ... ...}

\end{abstract}
\subsubsection*{Keywords}
Battery models, equivalent circuit modeling, joint parameter-state estimation, photovoltaic models, steady-state
\section*{Nomenclature}
% \tablecaption{Nomenclature for Solar PV and Battery Systems}
\begin{supertabular}{p{0.11\textwidth}p{0.28\textwidth}p{0.04\textwidth}}
    \label{tab:nomenclature}\\
    \multicolumn{3}{l}{\textbf{General Symbols}} \\ \hline\\
    ${\mathcal{P}}$              & Set of circuit model parameters & \\
    ${\mathcal{P}_\text{U}}$    & Set of unknown circuit model parameters & \\
    ${\mathcal{H}}$              & Set of combined network constraints & \\
    ${\mathcal{Z}}$              & Set of measurements & \\
    $\boldsymbol{\mathcal{W}}$              & Error covariance matrix & \\
    $\eta_{inv}$                            & Inverter efficiency (DC-AC) & \\
    $\eta_{rec}$                            & Rectifier efficiency (AC-DC) &  \\
    $\hat{\mathcal{P}}_\text{U}$                    & Parameter estimates & \\
    $N_s$                                   & Number of samples &\\
    $N_c$                                   & Number of components&\\
    \multicolumn{3}{l}{\textbf{Grid Specific Symbols}} \\ \hline\\
    ${\mathcal{N}}$              & Set of all grid buses & \\
    ${\mathcal{N}_{ZI}}$         & Set of zero-injection buses & \\
    ${\mathcal{M}}$              & Set of grid measurement circuits (e.g., RTUs measuring injection power) & \\
    ${\mathcal{G}}$              & Set of grid components without measurements (e.g., line, transformer, etc.) & \\
    $\Tilde{V}_k$                           & Complex nodal voltage at bus $k$ &  \\
    $V^r_k, V^i_k$                           & Real/imaginary voltage at bus $k$ &  \\
    $V^r_{kl}, V^i_{kl}$                     & Real/imaginary voltage difference between bus $k$ and $l$ &  \\
    $|V|_k, \theta_k$                       & Voltage magnitude/angle at bus $k$ &  \\
    $P_k, Q_k$                               & Active/reactive power injection at bus $k$ &  \\
    $\PGk{}, \QGk{}$                        & Real and reactive power injection of generator at bus $k$ &  \\
    $\PLk{}, \QLk{}$                        & Real and reactive power injection of load at bus $k$ &  \\
    $\Vmagk{}$                              & Generator regulated voltage magnitude at bus $k$ &  \\
    $\Gkz{}, \Bkz{}$                        & Measurement feature transformation terms at bus $k$ &  \\
    $P_{k,z}, Q_{k,z}$                       & RTU injection measurements of real/reactive power at bus $k$ &  \\
    $|V|_{k,z}$                              & RTU injection measurements of voltage magnitude at bus $k$ &  \\
    $\ItildG{}, \ItildL{}$                  & Generator and load complex current injection at bus $k$ &  \\
    $\nrtur{}, \nrtui{}$                    & Real and imaginary RTU measurement noise at bus $k$ &  \\
    $G_{kl}, B_{kl}$                         & Admittance terms between bus $k$ and $l$ &  \\
    $h^{Grid}_{k}$                           & Set of current-voltage constraints for grid network physics at bus $k$ & \\
    &&\\
    \multicolumn{3}{l}{\textbf{PV Specific Symbols}} \\ \hline\\
    ${\mathcal{K}}$              & Set of bulk grid-connected solar PV & \\
    $\RS{}, \RSH{}$                         & Series and shunt resistor of PV circuit &  \\
    $Z_{Load}$                               & Grid equivalent impedance seen by the solar PV &  \\
    $\VSH{}$                                & Voltage across shunt resistor in PV circuit, equal to open circuit voltage &  \\
    $\VPV{}, \zpvv{}$                       & Solar PV voltage at inverter terminal and its measurement &  \\
    $\VD{}$                                 & Diode voltage in solar PV circuit &  \\
    $\IPH{}, \zphi{}$                       & PV photocurrent generation and measurement & \\
    $\ID{}$                                 & Diode current in PV circuit &  \\
    $\Izero{}$                              & Saturation current of diode in PV circuit &  \\
    $\IPV{}, \zpvi{}$                       & Solar PV terminal current and its measurement &  \\
    $P_\text{PV}$                           & PV DC injection power to inverter &  \\
    $\npvv{}, \npvi{}, \nphi{}$            & Solar PV voltage, current output, and photocurrent measurement noise terms &  \\
    $w^\text{PV}_\text{I}, w^\text{PV}_\text{V}, w^\text{PV}_\text{PH}$ & Weights for current, voltage, and photocurrent measurements for PV &  \\
    $h^{PV}_{k}$                             & Current-voltage constraints for $k^{th}$ PV circuit on the grid & \\
    &&\\
    \multicolumn{3}{l}{\textbf{Battery Specific Symbols}} \\ \hline\\
    ${\mathcal{L}}$              & Set of battery systems connected to bulk grid & \\
    $\Vsoc{}$                               & Voltage representation of battery SOC, range between 0V to 1V for empty to full &  \\
    $\VOC{}$                                & Battery open circuit voltage &  \\
    $\VBatt{}, \zbattv{}$                   & Battery terminal voltage and its measurement &  \\
    $\IBatt{}, \zbatti{}$                   & DC battery current and its measurement &  \\
    $\Ccap{}$                               & Battery equivalent capacitor & \\
    $\RSE{}, \RSD{}$                        & Series and self-discharge resistor of battery &  \\
    $h^{\text{Bt}}_{k}$                     & Current-voltage constraints for the $k^{th}$ battery & \\
    $\nbattv{}, \nbatti{}$                  & Battery terminal voltage and current measurement noise terms &  \\
    $w^\text{Bt}_\text{I}, w^\text{Bt}_\text{V}$ & Weights for current and voltage measurements for battery & \\
    &&\\
\end{supertabular}

% \end{IEEEdescription}

\thanksto{\noindent This material is based upon work supported by the National Science Foundation through contract ECCS: 2330195 and the U.S. Department of Energy's Office of Energy Efficiency and Renewable Energy (EERE) under the Solar Energy Technologies Office Award Number DE-EE0010147. The views expressed herein do not necessarily represent the views of the U.S. Department of Energy or the United States Government.}

\section{Introduction}
%\subsection{Motivation of the research}\label{sec:Moti}
 
% \TODO{[Emphasize the impact of HES instead of describing ACSE]}\\

\noindent \textbf{Motivation:} Modern transmission networks now integrate emerging energy sources, such as utility-scale solar, battery storage, and fuel cells, alongside traditional synchronous generation. Among these, solar PV and battery systems are experiencing the fastest growth in several US regions \cite{Short-Term_Energy_Outlook_2024}, and this trend is expected to continue. Unlike traditional generators, these sources show i) higher fluctuation in power output, ii) a wider range of control with power electronics, and iii) increased instrumentation through more measurements.
In some areas, solar PV and battery resources now dominate daytime generation \cite{Bowers_Fasching_Antonio_2023}, providing essential services like carbon-free energy, peak shaving, and demand response. 
\textit{Therefore, an accurate real-time model and parameter state estimates are necessary to support these applications.}
% Considering the trend of moving to fully renewable, in this work we focus on HES composed of utility-scale solar farms and battery systems. 
%HES need to achieve the carbon footprint saving goal to compensate for the extra cost compared to traditional generation systems and maintain a reliable generation at the same time. Such a requirement can be achieved with the help of adequate control of the system \cite{gupta2011modelling}, which requires sufficient measurements and good estimation.
% HES can perform many grid services, including voltage regulation, frequency regulation, peak shaving, etc. We need accurate models for these grid components to provide an accurate state estimation to participate in various grid services.
However, gaps remain in the effectiveness of current estimation methods.
Traditional state estimation (SE) relies on constant power PV and PQ models, using power measurements from remote terminal units (RTUs) to estimate the states of photovoltaic (PV) and battery systems. This approach simplifies the solar PV and battery model to a power source or a negative load and neglects critical internal states.
%,while the internal state could be essential for us. 
% For example, an equivalent power source or a negative load approximates a utility-scale battery storage system, which overlooks internal states like the battery's terminal voltage and state of charge.
For example, the state of charge (SoC) of a battery is critical to estimate when planning for grid services like peak shaving or demand response, as SoC reflects the current capacity of the battery storage system. 
PQ and PV models limit us to the Coulomb counting method when it comes to SoC estimation. 
% it has some drawbacks: the initial state of SoC can not be calculated through this method, the error in current input accumulates during the counting.
%and the accuracy is affected by temperature, battery history, discharge current, and cycle life of the battery \cite {chang2013state}. 
A more detailed model can provide an estimated open circuit voltage ($V_\text{OC}$) to conduct open circuit voltage-based SoC estimation, thus allowing a hybrid estimation of SoC, that is, to estimate SoC with both the Coulomb counting method and open circuit voltage to calculate  SoC to find the solution that can satisfy both conditions. Literature like \cite{wang2007combined} has shown that the hybrid method generally provides better SoC estimation than individual methods.

Similarly, for utility-scale solar PV systems, constant power source approximation limits the potential to expand the model to account for weather conditions, including irradiance temperature, and incorporate additional measurements from the solar PV systems. 
The utility-scale solar PV systems are generally well-equipped with sensors and communication infrastructure. This means the state variables in a utility-scale solar PV system can be measured directly by power electronic sensors and meteorological sensors, making it usually observable \cite{fang2020state}. It is natural to make use of these measurements and enhance the redundancy level of our system.

% thus generating inaccurate and nonrealistic results and then further causing a misjudgment of its peak shaving ability.
% Aggregating PV and battery model parameters into constant power models also has drawbacks.
However, it requires molding and parameterization of solar PV and battery storage systems to include measurements from them. The parameters generally are from spec sheets or lab test results, which represent either a batch or a type of equipment but do not accurately reflect exact parameters for the specific asset connected to the grid at a given time and condition and these inaccurate parameters will deteriorate the accuracy of state estimation (SE).
\textit{Therefore, in this work, we focus on estimating the states of battery and solar-connected transmission networks with circuit-based detailed models for solar PV and battery components and an assumption that certain model parameters may be erroneous.} 
The approach falls under the broad paradigm of joint parameter-state estimation.

To achieve this goal we build on prior works on equivalent circuit modeling and estimation \cite{CircuitTheoretic2020}, \cite{sang2024circuit}, and \cite{jovicic2020enhanced}.
Both solar PV and battery systems can be characterized via equivalent circuit models.
Therefore, we include these circuit models within the broader equivalent circuit of the transmission network.
%While reviewing these works, we noticed that the equivalent circuit approach provides a generic framework to include solar, battery systems, and bulk grid.
We use the zeroth-order circuit model \cite{chen2006accurate} to represent battery steady-state behavior and the single-diode circuit model \cite{wang2008steady} for the solar PV system.
Further in our approach for joint parameter-state estimation, we consider additional measurements available from the battery and solar PV systems in addition to traditional remote terminal unit measurements.
%Similarly,  build equivalent circuit models of traditional power system components.
%Previous HES-related works like \cite{chen2006accurate}, \cite{wang2008steady}, and many other works show equivalent circuit models are touted as a standard approach to model new grid resources such as batteries and PV. 
%Works like \cite{jereminov2016improving} provide a power flow analysis using PV and battery as examples to show the merit of connecting equivalent circuit models of HES and traditional power system analysis. 
%Also, in our previous work \cite{sang2024circuit}, we modeled the grid as an equivalent circuit and conducted parameter state joint estimation (PSE) with the model by replacing measurement with an equivalent measurement circuit, utilizing the substitution theorem of the circuit. 
% After reviewing the battery and solar equivalent circuit model, 
%It is natural for us to combine the solar PV and battery circuit models and grid circuits by replacing the components in the transmission system with the equivalent circuits. By this measure, we extend our grid model to solar-battery-augmented grid models. This extension can help parameterize and estimate these models with the \textit{new} and standard RTU/PMU measurements under one framework. 
% Also, the circuit formulation allows us to use and include the existing module in our joint PSE formulation.

\noindent \textbf{State-of-art and limitations:} 
% PSE is not a new problem.
Most current research on joint parameter-state estimation [4] and [10] focuses on transmission or distribution networks, with battery and solar PV systems often modeled like synchronous generation or negative load. 
Alternatively, they estimate [11]-[16] stand-alone battery and solar PV systems without coupling them to the grid.

The research is limited when considering the estimation and parameterization of combined solar PV, battery, and grid systems. 
For instance, \cite{fang2020active} estimates PV fitting function parameters separately before applying them to the traditional SE formulation, while \cite{fang2020state} introduces an augmented state estimation (ASE) approach using multiple measurements. Both works focus on estimating specific PV fitting function parameters rather than generic PV models.

Stand-alone PV system estimation research, such as \cite{easwara1986nonlinear}, \cite{oliva2014parameter}, and \cite{alhajri2012optimal} solves optimization problems to estimate the PV states and parameters by minimizing the norm of the current residual to fit empirically obtained I-V curves.
Stand-alone battery parameter estimation works \cite{miniguano2020genral}, \cite{rahimi2013online} minimize the residual norm between experimental measurements and model estimate.  
Other works like \cite{jusoh2020accurate}, minimize the time step integration of mismatch between voltage estimates and spec sheet voltage.
These are all stand-alone techniques and do not include traditional grid measurements (e.g., from remoter terminal units [RTUs]) to improve their estimation accuracy.
%Solar PV, battery storage, and the grid should follow energy conservation laws (power balance) across systems. However, separate estimations often lead to inconsistencies at the interconnection points, resulting in discrepancies between the models. These inconsistencies can complicate decision-making and hinder effective coordination between the two systems.

% These heuristic methods minimize the mismatch to a measurement (terminal voltage or current output) in their objective under some physical constraints. The method aims to find the closest possible parameter under physical constraints that can minimize the residual, effective when a good enough measurement is provided. These measurements can be accurate for a single device, but for utility-scale equipment connected to the grid, the ground truth of a single tested device may not be a good representation of the whole system, thus decreasing the reliability of these methods.
% \TODO{For now there is limited material about co-estimating of HES and TG, should I review the separated methods and keep them here?}
% Due to differences in measurement equipment and problem formulation, solar PV, battery, and power system components will likely produce different estimations for the same state or parameter. Regardless of the magnitude, inconsistencies persist when these components are estimated separately.

We find that key limitations of the state-of-the-art include the i) inability to represent the detailed physics of grid-tied battery and PV systems, ii) inability to include a wide array of measurements, and iii) inconsistent estimations of the states at the boundary of battery and PV system and the grid.

% \TODO{(I do not have a reference example now though it is good to have one)} 

\noindent \textbf{Insight of the problem:}
% \TODO{What is one or two points that are so special that make ECF helpful here?}
%To address the gap in the previous section, we propose a parameter-state estimation (PSE) framework that integrates the grid with measurements from utility-scale solar PV and battery systems connected behind the bus.
We will address these gaps by building a circuit-theoretic joint parameter estimation framework for transmission grid networks, including detailed models for utility solar PV and battery systems.
In this approach, first, we will develop an aggregated circuit to model the physics of combined grid, utility-scale PV, and battery components.
Next, in the aggregated circuit, we will replace or substitute the circuit element with its measurement equivalent where measurements are available.
%We expect replacing separated models with one aggregated model can solve the mentioned limitations. Thus, we will first build circuit models for each component Then, we include available measurements from each component of the circuit. %including terminal voltage and current. Additionally, photocurrent ($I_{PH}$) proportional to irradiance is represented as measurement ($I_{PH,z}$) for solar PV, all measurements have a corresponding noise ($\eta$).
 %and help us decrease the effect of noise. provide us with better estimation (assume unbiased measurements). 
Finally, we will solve a large-scale optimization problem to obtain the system estimates. 
We will constrain the optimization problem by the physics of the measurement-based aggregated circuit, and in the objective, we will minimize noise terms in the measurement circuits.
We will also model any potentially erroneous parameters as unknowns. 
%We will show that this approach provides a more accurate representation of the system and is robust against erroneous parameters.

% This research aims to provide a PSE framework to incorporate the grid and measurements from utility-scale solar PV and battery systems behind the bus. So that for the grid side more information on the system can be included in the formulation, for the HES side the grid side constraints can further limit the feasible region.
% For the grid portion of the problem, the technique leverages additional measurements from behind the bus. For the solar PV and battery storage systems, the method defines a feasible region based on the physical constraints of the transmission grid.

We anticipate several benefits from the proposed method in overcoming the mentioned limitations:
\begin{itemize}
    \item \textbf{Generality to measurements}: The approach includes and uses a heterogeneous set of measurements from the traditional grid meters and utility-scale battery and PV systems.
    \item \textbf{Solution consistency}: The estimates are consistent at the individual systems interconnect boundaries.
    \item \textbf{Robust against erroneous parameters}: The approach is robust to erroneous parameters in the system.
\end{itemize}

% For example, DC-AC inverters used for solar PV systems can provide us with power output data and voltage information, and the on-field weather station can also provide temperature and solar irradiance measurements. By incorporating more measurements, this framework is expected to provide better estimation.

% In all, this work focuses on tackling limitations by the following measures:
% \begin{itemize}
%     \item[1.] Incorporate solar and battery system measurements behind the bus to utilize more available measurements and deliver better estimation results compared to the traditional method
%     \item[2.] A co-estimation between the grid, PV, and battery system will provide a unified estimation of the state.
%     \item[3.] A PSE framework that combines the grid, PV, and battery system can shrink the feasible region of their variables.
% \end{itemize}

% \TODO{Is expanding it to incorporate HES measurements enough? Then, the scalability will be needed to tell the story, as the ability to integrate many subcircuits is required if using the circuit to describe a utility-scale solar farm}

\section{Preliminaries} \label{modeling}
\noindent To build the joint circuit-based parameter-state estimation algorithm with grid-tied battery and solar PV systems, we describe the equivalent circuit models for grid, battery, and PV systems and state-of-the-art in circuit-theoretic state estimation.

\subsection{Equivalent Circuit Modeling}\label{sec:ECM}
In equivalent circuit modeling (ECM), we build equivalent circuits for individual grid components and aggregate them to represent the overall grid physics.

\subsubsection{Grid Modeling}
\begin{figure*}[ht]
  \includegraphics[scale=0.07]{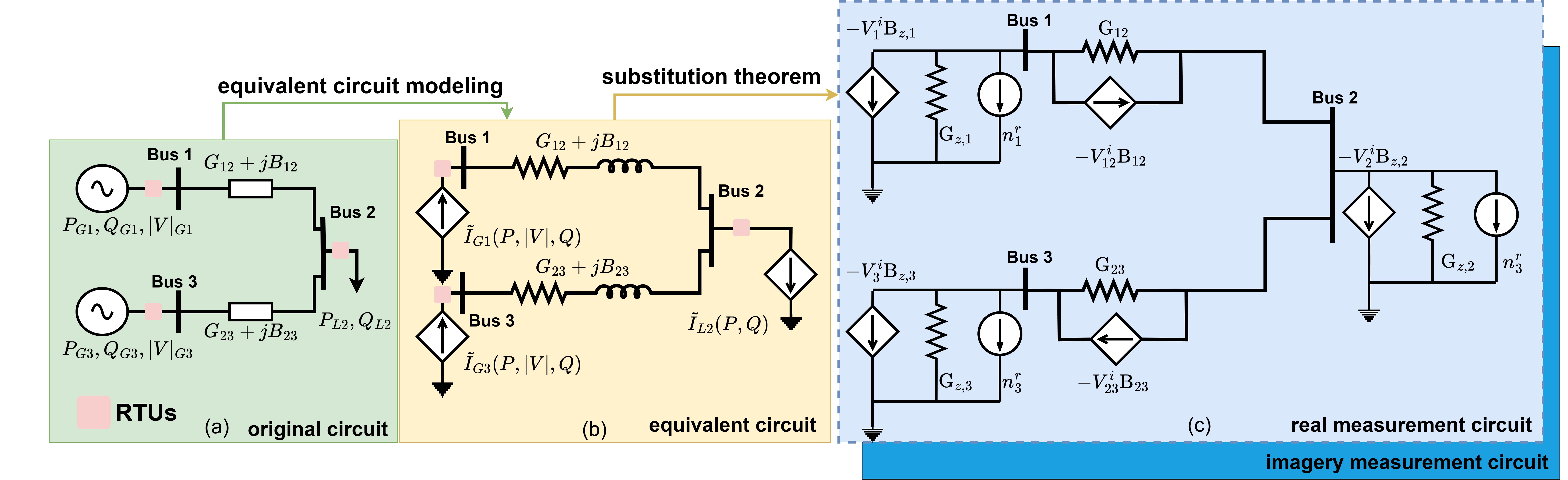}
  \caption{(a)3-bus circuit example (b)with equivalent current source account for power injection (c)with RTUs represented as measurement circuits.}\label{fig:3bus_circuit}
\end{figure*}

In ECM, circuit models for generator (PV), load (PQ), branch, transformer, and other grid components are derived from KCL-based current-voltage relationships. 
For example, consider node 2 in the 3-bus system in Fig.\ref{fig:3bus_circuit}
%For bus 2, the ’PQV’ formulation power balance is as follows: 
%\begin{align}
%    &P^2_{inj} = P^{21}_{flow}(V_1,V_2) + P^{23}_{flow}(V_2,V_3)\\
%    &Q^2_{inj} = Q^{21}_{flow}(V_1,V_2) + Q^{23}_{flow}(V_2,V_3)
%\end{align}
where the Kirchhoff current law (KCL) describes the nodal IV relationship: 
\begin{align} 
    &I^r_{L,2} + V^r_{21}G_{21} - V^i_{21}B_{21} + V^r_{23}G_{23} - V^i_{23}B_{23} = 0 \label{load_currentR}\\
    &I^i_{L,2} + V^i_{21}G_{21} + V^r_{21}B_{21} + V^i_{23}G_{23} + V^r_{23}B_{23} = 0 \label{load_currentI}
    % &- I^r_{L,2} + I^{r}_{flow,21}+ I^{r}_{flow,23} = 0\\
    % &- I^i_{L,2} + I^{i}_{flow,21} + I^{i}_{flow,23} = 0
\end{align}
The real and imaginary load currents in \eqref{load_currentR} and \eqref{load_currentI}, are nonlinear functions of real and reactive power and complex voltages:
%Substitution theorem allows replacing of known branches with branches that make the same voltage and current, thus a given power source that injects $\text{P}_{G1},\text{Q}_{G1},\text{at bus voltage magnitude } |\text{V}|_{G1}$ can be replaced by a current source $\Tilde{\text{I}}_{G1}$, then it can be split into real and imaginary circuit:
\begin{align}
    &\text{I}^r_{L,2} = \frac{P_{2}V^r_{2} + Q_{2}V^i_{2}}{(V^r_{2})^2+(V^i_{2})^2}\\
    &\text {I}^i_{L,2} = \frac{P_{2}V^i_{2} - Q_{2}V^r_{2}}{(V^r_{2})^2+(V^i_{2})^2}
\end{align}

These relationships (and other similar relationships) can be directly mapped to circuit models. 
For instance, see the ECM representation of the power grid model in part (a) of 
Fig. \ref{fig:3bus_circuit} in part (b) of Fig. \ref{fig:3bus_circuit}.
Refer to \cite{Bromberg2015ECF}, \cite{jereminov2016improving}, and \cite{pandey2018robust}  for a comprehensive description of ECM for traditional grid components.

\subsubsection{Solar PV modeling}
\begin{figure}
    \centering
    \includegraphics[scale = 0.09]{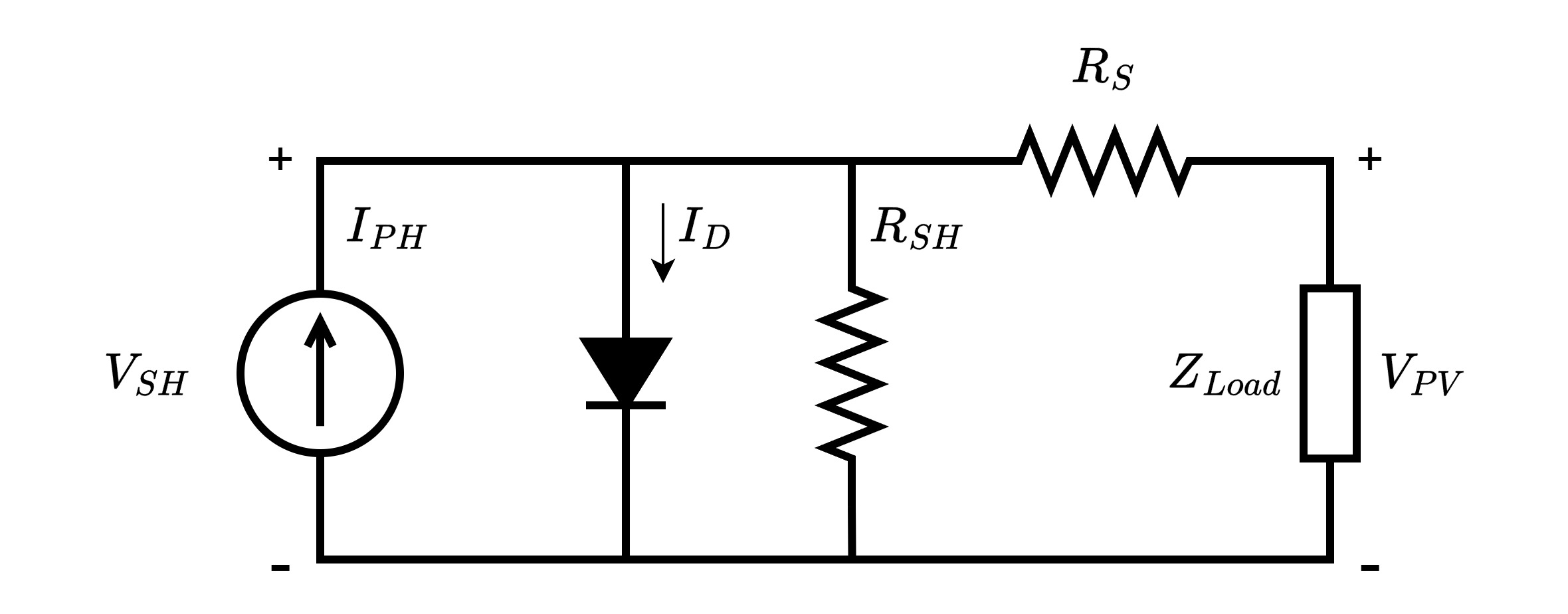}
    \caption{Equivalent circuit of single diode model.}
    \label{fig:PV_ECF}
\end{figure}

To represent solar PV systems in ECM, we use the circuit-based single diode model (SDM) \cite{humada2016solar} shown in Fig. \ref{fig:PV_ECF}. 
SDM is widely used for its simplicity \cite{ikegami2001estimation, naeijian2021parameter, xiao2012efficient} and is sufficient to represent the amorphous silicon PV system \cite{gow1999development} in comparison to more detailed double diode model (DDM). 
% equivalent circuit of the single diode PV model is shown in . 
% The model's circuit representation with the pre-fixed choice of representation control mode can be found in\cite{jereminov2016improving},\cite{cubas2014analytical}.

In SDM, the photovoltaic, when not illuminated, behaves like a semiconductor diode whose voltage-current relationship is given by: 
\begin{equation}
    \ID{} = \Izero{}\left[exp\left(\frac{\text{q}\VD{}}{\text{nk}T}\right) - 1\right]\label{eq:shockley}
\end{equation}
\noindent where the diode current ($\ID{}$) is calculated with Shockley diode equation \eqref{eq:shockley}. 
The voltage across the diode is noted as $\VD{}$. 
The k is the Boltzmann constant, and q is the electric charge for a single electron ($1.602e-19\;\text{C}$). 
$\Izero{}$ represents the reverse saturation current of the diode, and $T$ is the cell temperature in Kelvin. 
$\text{n}$ is the ideality factor of the diode, it is a factor that describes the diode quality and material of the diode (usually between 1 and 2). 
When $\text{n}$ is unity, it represents an ideal diode; however, including leakage current pushes ideality factor $\text{n}$ away from unity.

When illuminated, the diode provides a photocurrent ($\IPH{}$), modeled in SDM via a current source whose magnitude is a function of irradiance.
This current divides into the diode current ($\ID{}$), leakage current ($\ISH{}$) through shunt resistance $\RSH{}$, and output current ($\IPV{}$). 
The losses in the interconnection junction box are modeled via series resistance ($\RS{}$).
%flows into the observed grid load. 
%To account for the current-voltage relationship, more components are included in the circuit: A current source describing the current produced by photon-electron transfer in solar panels ($I_{\text{PH}}$), a series connected resistor ($R_S$) to the current source for $I_\text{S}$ and a shunt resistor added in parallel to the current source and diode ($R_{SH}$) for ($I_\text{SH}$), an impedance $Z_{\text{load}}$ account for ($I_\text{PV}$).
% The single-diode model includes the following components:
% \begin{itemize}
%     \item a current source describing the current produced by photon-electron transfer in solar panels. The semiconductor material, the cell temperature, and the solar irradiance upon the cell decide this term.
%     \item a diode that is connected in parallel to the current source, it accounts for the p-n junction material effects
%     \item a series connected resistor ($\text{R}_S$) to the current source that models the loss in cell solder bounds, interconnection, junction box, etc.
%     \item a shunt resistor added in parallel with the current source and diode ($\text{R}_{SH}$) models loss in the current leakage
% \item the inverter loads the PV array with impedance $\text{Z}_{Load}$
% \end{itemize}
We can extend the SDM model to include PV system controls without loss of generality. 
For instance, \cite{jereminov2016improving} and \cite{cubas2014analytical} introduce circuit representation for maximum power point tracking (MPPT) and volt-var control, respectively.
%Note that different inverter control schemes load PV arrays differently by changing the impedance seen by the PV. Take maximum power point Tracking (MPPT) [xx]
%In \cite{jereminov2016improving}, to model MPPT control, the nonlinear impedance $Z_{\text{load}}$ is adjusted implicitly to maximize output from the solar PV system by sweeping $V_{\text{PV}}$. Consider the $P_{\text{PV}}-V_{\text{PV}}$ curve has a monotonic decreasing first derivative of $P_{\text{PV}}$ by $V_{\text{PV}}$, when the first derivative reaches zero, we land at the maximum of $P_{\text{PV}}-V_{\text{PV}}$ curve\cite{xiao2017photovoltaic}.
%Mathematically, it implies that: 
%\begin{equation}
%    \frac{\partial(P_{\text{PV}})}{\partial V_{\text{PV}}} = \frac{\partial(I_{\text{PV}}V_{\text{PV}})}{\partial V_{\text{PV}}
%    } = I_{\text{PV}} + V_{\text{PV}}\frac{\partial I_{\text{PV}}}{\partial V_{\text{PV}}} = 0
%\end{equation}
%

%By KCL, the equivalent circuit of the PV indicates that: 
%\begin{equation}
%    -I_{\text{PH}} + I_\text{{D}} + I_\text{{SH}} + I_\text{{S}} = 0 
%\end{equation}
%and:
%\begin{equation}
%    I_{\text{PV}} - I_\text{{S}} = 0
%\end{equation}
%The current of shunt resistance ($\text{R}_{SH}$) is calculated by:
%\begin{equation}
%    I_\text{{SH}} = \frac{V_{\text{OC}}}{\text{R}_{\text{SH}}} = \frac{V_{\text{PV}} + I_S\text{R}_\text{S}}{\text{R}_{\text{SH}}} 
%\end{equation}
%% \TODO{[Unit of q is $eV$? make sure the units are correct, and mention how do you get the parameters and whether they are a good representative of the system]}
%\TODO{Add linearized circuit?}
%
\subsubsection{Battery modeling}
\begin{figure}
    \centering
    \includegraphics[scale = 0.1]{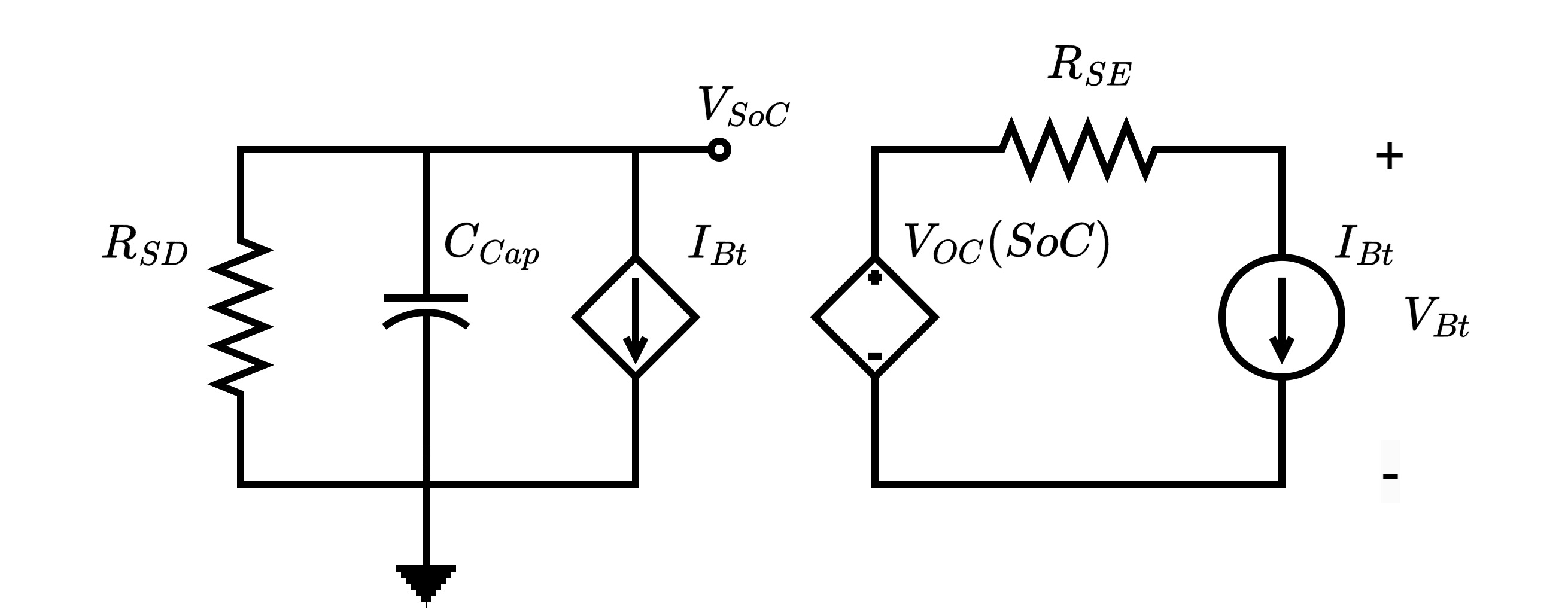}
    \caption{Equivalent circuit of $0^{th}$ order battery model.}
    \label{fig:batt_ECF}
\end{figure}

\noindent Equivalent circuits can also model the voltage-current characteristics of a battery. %Like other grid models include photovoltaic systems, these fit into the aggregated circuit representation.
% The equivalent circuit models of batteries have different \textit{orders} which indicate models' number of \TODO{[Explain the order of battery model here or no]}. For battery dynamics in a fast system, a higher-order model is favored to capture its behavior. 
As the estimation and parameterization techniques in this paper focus on the \textit{near} steady-state response of the aggregated grid, we neglect current-voltage transient behavior during charging and discharging and use a zeroth-order approximation of the battery circuit model in \cite{chen2006accurate} to model voltage-current characteristics. 
The zeroth-order circuit is shown on the right of Fig. \ref{fig:batt_ECF}. 
The left of the figure represents the battery lifetime or the state of charge (SoC) dynamics.
The battery lifetime subcircuit has the following components: a capacitor ($\Ccap{}$) that is analogous to the rated capacity of the battery at full charge, a controlled current source that represents the current output ($\IBatt{}$), and a resistor ($\RSD{}$) connected in parallel to account for the battery's self-discharge.

The voltage across the capacitor ($\Vsoc{}$) is analogous to the SoC of the battery and ranges between 0V and 1V, representing the fully discharged and fully charged state of the battery, respectively. 
The capacitance $\Ccap{}$ represents the ratio of the total charge stored in the battery as a function of  $\Vsoc{}$:
\begin{equation}
    q_{\text{Cap}} = \Ccap{}\Vsoc{}
\end{equation}

\noindent With capacitor voltage $\Vsoc{}$, and stored charge $q_{\text{Cap}}$, the rate of charge of $q_{\text{Cap}}$ is given by:
\begin{equation}
    I_{\text{C}} = \Ccap{}\frac{d\Vsoc{}}{dt}
\end{equation}
In the integral form, it can be written as:
\begin{equation}
    V_{\text{SoC},t} - V_{\text{SoC},0}= \frac{1}{\Ccap{}}\int_0^t I_\text{C} (t) dt \label{func:Vsoc}
\end{equation}
and the total charge transfer for time ranging from $t \rightarrow\tau$ in terms of ampere-hours is given by:
\begin{equation}
    q_{\text{Cap},t} - q_{\text{Cap},0} = \int_0^\tau I_\text{C} (t) dt = -\int_0^\tau \IBatt{} (t) dt\label{func:qt}
\end{equation}
The KCL-based relationship between the battery current $I_{Bt}$, self-discharge current ($I_\text{SD}$), capacitor current ($I_\text{C}$):
\begin{equation} \label{func:batt_leftCKT}
    I_{\text{SD}} + I_\text{C} + \IBatt{} = 0
\end{equation}
%As $R_{SD}$ is very large, we can the approximate the equation with:
%\begin{equation}
%    I_{\text{C}} + \IBatt{} = 0
%\end{equation}

%When we combine equation \eqref{func:Vsoc} and \eqref{func:qt} a new function showing the relationship between total charge and $\Ccap{}$ can be summarised, that is, when we assume $\Vsoc{}$ = 1V when fully charged by $q_{full}$ charges, the capacity of the equivalent capacitor should be:
%begin{equation}
%    \Ccap{} = \frac{q_\text{full}}{V_\text{SoC,full}} = \frac{q_\text{full}\text{(Coulomb)}}{1\text{(Volt)}}
%\end{equation}

\noindent By representing $I_{\text{C}}$ and $I_{\text{SD}}$ as functions of $V_{\text{SoC}}$ we get the final form in (\ref{func:batt_leftCKT1}):
\begin{equation} \label{func:batt_leftCKT1}
    \frac{\Vsoc{}}{\RSD{}} + \Ccap{}\frac{d\Vsoc{}}{dt} + \IBatt{} = 0
\end{equation}

The right half of Fig.\ref{fig:batt_ECF} depicts the voltage-current characteristics of the battery. 
It has the following components: a voltage-controlled voltage source ($\VOC{}$), a series resistor ($\RSE{}$), and a controlled current source to represent current output ($\IBatt{}$) during constant power charge/discharge.
We model the open circuit voltage as a fitted linear function of SoC ($\VOC{}$) with two fitted parameters $a$ and $b$:
\begin{equation}
    \Vsoc{} = a + b\VOC{} 
\end{equation}
%The internal resistance $\RSE{}$ is connected in series with the voltage source. 
The terminal voltage of the battery ($\VBatt{}$) is given by Kirchhoff voltage law (KVL):
\begin{equation}
    -\VOC{} + \IBatt{}\RSE{} + \VBatt{} = 0 
\end{equation}

\subsubsection{Inverter modeling} \label{sec:inverter_modeling}
The power converter bridges the DC battery and solar PV systems to the AC grid. 
The power transfer efficiency between the AC and DC systems is a function of operating power.
For the estimation problem, we model the inverter efficiency with curve-fitted efficiency functions.
% We assume a maximum power point tracking (MPPT) control strategy is applied to force the solar array to operate at its maximum power point ($P_\text{MPP}$). 
We also assume the unity power factor for solar PV systems operation \cite{jereminov2016improving}, thus for a PV at bus $k$:
\begin{align} \label{eq:inv_eff}
    P_{{k}} &=\eta_{inv} P_{\text{PV}}\\
    Q_{{k}} &= Q_{\text{PV}}= 0 
\end{align}

\noindent In \eqref{eq:inv_eff}, the inverter efficiency is a function of solar PV power output, $\eta_{inv} = f^D(P_{\text{PV}})$ \cite{gilman2015sam}. 

\begin{figure}
    \centering
    \includegraphics[scale = 0.26]{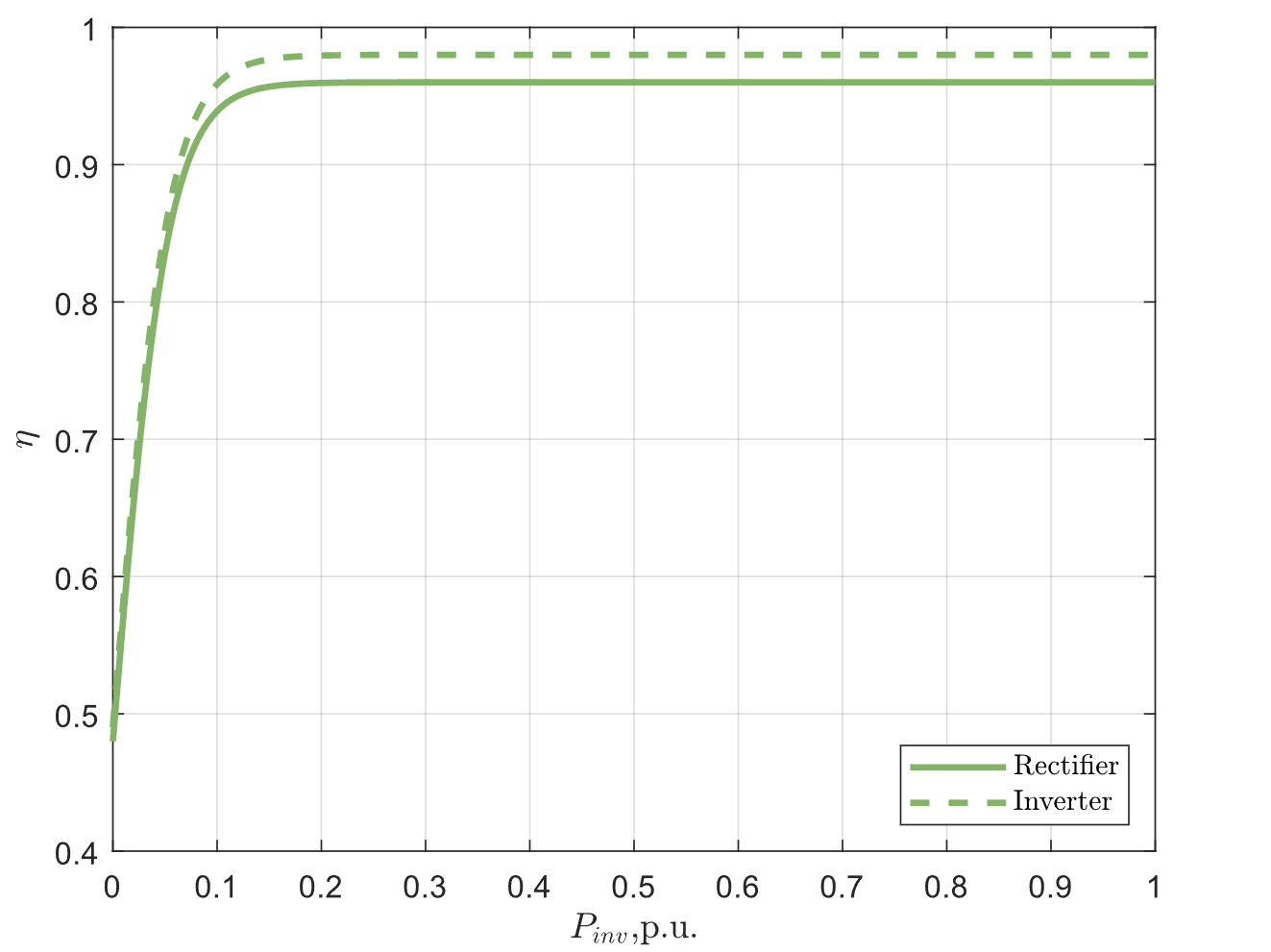}
    \caption{Sigmoid efficiency curve used in this work.}
    \label{fig:sigmoid}
\end{figure}

For this paper, akin to PV systems, we assume a unity power factor control for battery systems during discharge and inverter efficiency is given by $\eta_{inv} = f^D(P_{\text{Bt}})$.
We assume the maximum charging efficiency is lower than the discharging efficiency for the battery per inverter spec sheet. Therefore, for charging efficiency, we use a separate fitted function $\eta_{rec} = f^C(P_{AC})$.
For a battery connecting bus $k$, the equation is $\eta_{rec} = f^C(P_{k})$.
For both charging and discharging and inversion for PV systems, we chose the truncated sigmoid function to fit inverter curves for efficiency functions \cite{gilman2015sam}. The function has following form:
\begin{equation}
    \eta_{inv/rec} = \sigma(P) = \frac{M}{1 + e^{-\gamma P}}
\end{equation}
During curve fitting to mimic the efficiency function, we estimate parameters $M$ and $\gamma$ with data from the inverter datasheet. $P$ is positive in this setup thus we are using the positive part of the sigmoid function as shown in Fig. \ref{fig:sigmoid}.

% \subsection{substitution theorem}
% \TODO{Maybe it is quite a common knowledge that I do not need to explain}
\subsection{Circuit-theoretic State Estimation ($\cktse{}$)}\label{sec:grid ECF}
%  Builds on \textit{foundational} equivalent circuit-based approaches for ACSE \cite{CircuitTheoretic2020, jovicic2020enhanced} that solve physics-based Maximum Likelihood Estimation (MLE) and are compatible with both RTUs and PMUs measurements. MLE is an observation-
% driven method. It takes observed measurements as apriori
% information to estimate realizations of random variables from
% assumed distributions.
In this paper, we build a method to estimate the grid states (including battery SOC and solar output) with many utility-sized battery and PV systems.
We build the foundation of our work on existing circuit-theoretic grid estimation techniques in \cite{CircuitTheoretic2020}, \cite{jovicic2020enhanced}, which 
introduce
a convex circuit-based AC state estimation, termed \cktse{}.
\cktse{} defines relationships between network physics and measurements via affine constraints in the current-voltage (IV) framework. 
As an initial step, in \cktse{}, we replace any equivalent circuit downstream of a measurement with a measurement circuit following the substitution theorem.
However, as grid measurements are noisy, measurement circuits must include noise terms to satisfy KCL and KVL.
Finally, to estimate grid states, \cktse{} minimizes the norm of the noise terms in the measurement circuit subject to KCL and KVL-based grid constraints.
We illustrate the idea of measurement circuits in Fig. \ref{fig:3bus_circuit} wherein the transition from Fig. \ref{fig:3bus_circuit}(b) to Fig. \ref{fig:3bus_circuit}(c) shows the use of measurement circuits.
For instance, consider the node 2 load bus again in the 3-bus example.
The bus has a remote terminal unit (RTU), which measures the real and reactive power consumption of the load $(P_{z,2}, Q_{z,2})$ and voltage magnitude $(|V|_{z,2})$ at the node. The nonlinear relationships between power and voltage-current behavior are non-affine, thus causing limitations like convergence issues.
%on each bus that provide us with voltage magnitude, and real and imaginary power injection on each bus, can expands to incorporate measurements. 
%Again use the simple 3-bus example, the RTUs provide voltage magnitude ($|\text{V}|_\text{z}$), real and reactive power injection ($\text{P}_\text{z},\text{Q}_\text{z}$) of the measured bus.

To establish affine AC relationships and construct the measurement circuit for RTU measuring load parameters, these works (\cite{CircuitTheoretic2020}, \cite{jovicic2020enhanced}, \cite{jereminov2016improving}, and \cite{sang2024circuit}) rely on feature transformation \eqref{eq:feature_trans} that converts power and voltage measurements  $P_z$, $Q_z$, and ${|V|_z}$ into conductance and susceptance ($G_z$, $B_z$) circuit elements. This feature transform can reduce three-dimensional measurements to two dimensions:
\begin{align}\label{eq:feature_trans}
   G_{z,2} = \frac{P_{z,2}}{{\left({|V|_{z,2}}\right)}^2};\;\; B_{z,2} =\frac{Q_{z,2}}{{\left({|V|_{z,2}}\right)}^2}
\end{align}

\noindent With the transformation, the nodal network constraints (or the measurement functions) at each bus are affine. 
For example, the injection currents $I_{L2}$ at bus 2 are affine functions of unknown voltages:
\begin{align} \label{eq:rtu_injection_1}
    I_{L2}^{r} = G_{z,2} V_{2}^{r} - B_{z,2} V_{2}^{i} + n^{r}_2\\
    I_{L2}^{i} = G_{z,2} V_{2}^{i} + B_{z,2} V_{2}^{r} + n^{i}_2 \label{eq:rtu_injection_2}
\end{align}
\noindent

In addition to the transformed measurements, equation \eqref{eq:rtu_injection_1} and \eqref{eq:rtu_injection_2} also include noise terms ($n^{r}_2$ and $n^{i}_2$) to account for errors in power and voltage measurements. 
% The substitution theorem allows us to replace all equipment behind RTUs with measurement circuits of RTUs. Each injection bus satisfies this set of KCL equitations in the real and imaginary domains, as shown in Fig. \ref{fig:3bus_circuit}(c) which accounts for the real part equation \eqref{eq:rtu_injection_1}. The imaginary part has a similar structure which accounts for \eqref{eq:rtu_injection_2}.

%This method utilizes the "V-I" formulation instead of the traditional "PQV" formulation for linearity. The state variables are real $V_r$ and imaginary $V_i$ voltages instead of the voltage magnitude ($|V|$) and the angle $\theta$. 
Similarly, the approach also models other grid components with equivalent circuits.
The works \cite{CircuitTheoretic2020}, \cite{jovicic2020enhanced} describe the creation of these circuits in length. Subsequently, it applies the substitution theorem to replace any measured circuit elements with a measurement circuit, as shown in Fig. \ref{fig:3bus_circuit}(c).
% For example, consider the 3-bus system in Fig.\ref{fig:3bus_circuit}, where any component downstream of an RTU is replaced via a measurement circuit as shown in Fig.\ref{fig:RTU}, for the real part and imaginary part, the structure is similar.
%The substitution theorem is then applied to replace circuit components measured by RTU.
The approach then estimates the grid states by minimizing the norm of the noise terms in the measurement circuits subject to the constraints enforced by KCL equations at all buses $h^{grid}(V,n)$.
\cite{sang2024circuit} further extends this framework to account for certain unknown parameters.
% Consider a 2-bus example for a simple illustration of circuit-based state estimation as shown in Fig.\ref{fig:2bus_example}. 
% In this framework, all non-zero injection buses are assumed to be measured by remote terminal units (RTUs). The RTUs can provide voltage magnitude ($|V|$), real and reactive power injection ($P,Q$) of the measured bus. In this work, we note all measurements as $z$. For example, we note the measurements of bus $k$ as $z_{|\text{V}|}^k,z_{\text{P}}^k,$ and $z_{\text{Q}}^k$.

%This approach is built on certain relaxations. 
%To construct the measurement circuit for RTU nodal power injection, it applies a feature transform to reduce three-dimensional measurements to two dimensions:

%\begin{align}
%    G_z = \frac{P_z}{{\left({|V|_z}\right)}^2};\;\; B_z =\frac{Q_z}{{\left({|V|_z}\right)}^2}
%\end{align}
%\noindent With the transformation, the nodal network constraints (or the measurement functions) at bus $k$ and $l$ become affine, as the injection currents $I_{b}$ at bus $b \in \{k,l\}$ are linear functions of unknown voltages per KCL:
% \begin{align} \label{eq:rtu_injection_1}
%     I_{b,r} = G_{b,z} V_{b,r} - B_{b,z} V_{b,i} + \eta_{b,r}\\
%     I_{b,i} = G_{b,z} V_{b,i} + B_{b,z} V_{b,r} + \eta_{b,i} \label{eq:rtu_injection_2}
% \end{align}
%\noindent Both equations include noise terms $\eta$  accounting for imperfections in real and imaginary current measurements. Separately noted by $\eta_r$ and $\eta_i$. 
%%For each injection node, this set of KCL equations is satisfied.

While robust and provably convex, this method does not include detailed circuit-based models for battery or solar PV systems or incorporate their corresponding measurements. 
The current approach is limited to RTU and PMU measurements only.
The method also does not account for erroneous parameters.
\cite{sang2024circuit} considers erroneous parameters but is still limited to traditional grid parameters like line conductance and susceptance.
% \TODO{Do I need a section to introduce ECF like I had in PSCC that covers everything?}

\section{Circuit-theoretic Estimation of Combined Grid, Utility-scale PV and Battery Systems}
% \TODO{1.Construct optimization problem here}\\
% \TODO{2. Describe how it DC is normalized}

% \TODO{How is the estimation problem formulated by each component}
% \TODO{And how they meet each other in the optimization problem}
%By representing the battery and solar PV system as sub-circuits we expand the ECF SE framework to include the additional information from HES measurements.
\noindent Section \ref{sec:ECM} discussed the equivalent circuit models for grid, battery, and PV systems.
Section \ref{sec:grid ECF} discussed the circuit-theoretic state-estimation approach, which includes measurements within the grid circuit model.
Section \ref{sec:meas_ckt} will build measurement circuits for grid-tied battery and PV systems and, subsequently,
Section \ref{sec:model_assem} will formulate the circuit-theoretic estimation problem $\cktser{}$ for bulk grid components, including battery and PV systems, as an optimization problem.
The physics (defined by Kirchhoff's Laws) of the aggregated grid circuit, including measurement circuits, will define the constraint set of the 
$\cktser{}$ optimization problem.
The minimization of measurement noise terms in the measurement circuits will serve as the objective of the estimation problem.

%To solve the estimation problem, we will aggregate all network circuits, including measurement circuits and minimize the norm of all weighted noise subject to the KCL and KVL constraints of the aggregated circuit. 
%The  and show that the estimator is robust against erroneous parameters.

\subsection{Measurement Circuits for PV and Battery} \label{sec:meas_ckt}
% \TODO{What is measured in PV and Battery}
% \TODO{What is the problem (state and parameter) we are solving and}
% \TODO{What question we are answering (I. is the estimation better II. to level it increase the complexity}
% \TODO{Descirbe why in this research the ECF formulation is used as a comparison instead of traditional PQV formulation}

\noindent Section \ref{sec:grid ECF} (also \cite{jovicic2020enhanced} and \cite{CircuitTheoretic2020}) describes the inclusion of RTU and PMU measurements into circuit-based estimation framework.
%and prior works like \cite{jovicic2020enhanced} and \cite{CircuitTheoretic2020} formulate the optimization problem. 
Here, we derive circuits and corresponding equations for the inclusion of solar PV and battery measurements in the circuit-based framework. 
\subsubsection{Solar PV measurement circuits}
 \begin{figure}
    \centering
    \includegraphics[scale = 0.08]{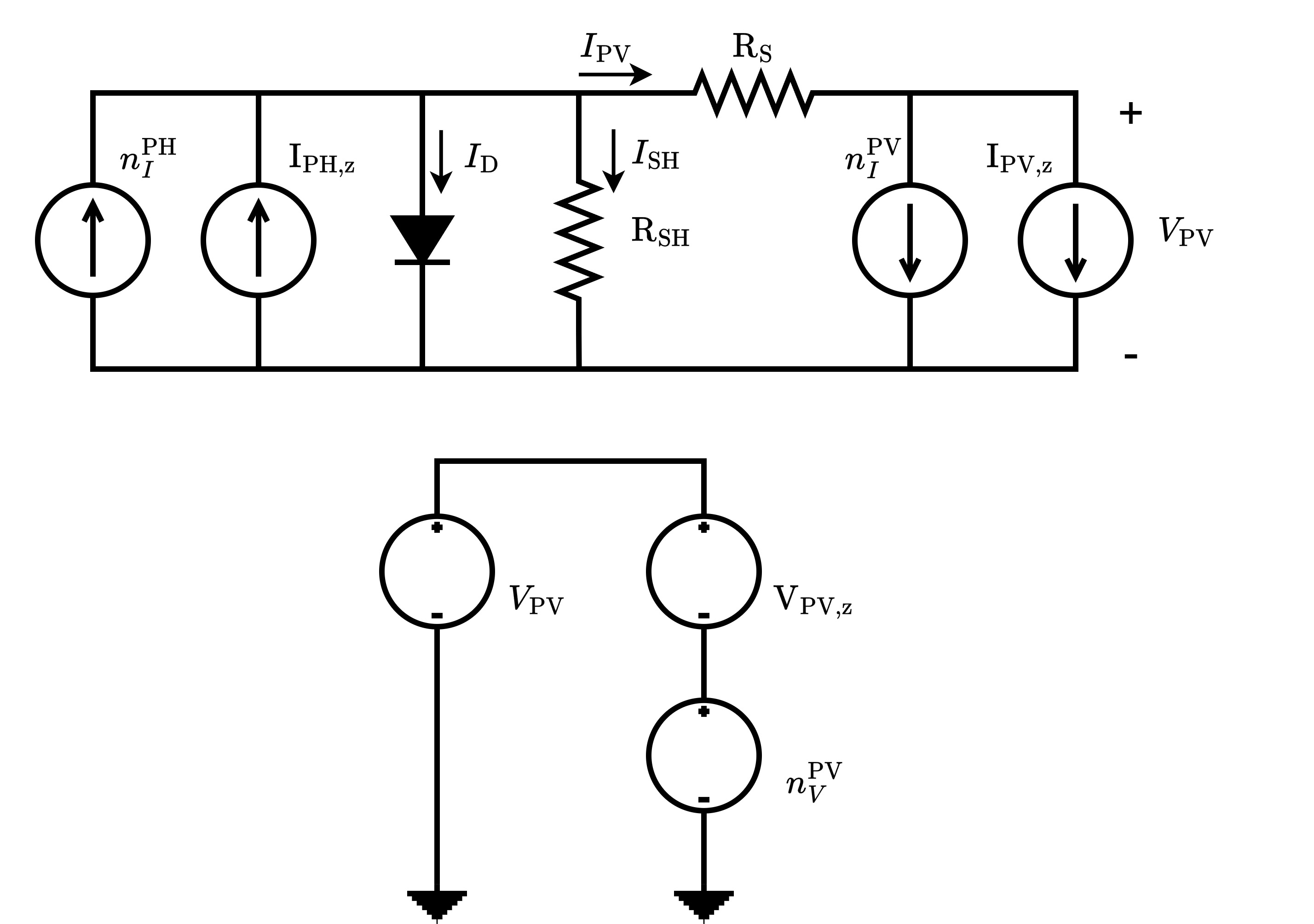}
    \caption{Equivalent circuit of the PV system including measurements.}
    \label{fig:PV_MEA_ckt}
\end{figure}
% \TODO{Explain how you get these measurements, $I_{PH}$ is explicit after the temp and irradiance is acquired. Anyways, explain clearly what measurements can be acquired}

\noindent We consider the following measurements for the solar PV systems: DC current ($\zpvi{}$) and DC voltage ($\zpvv{}$) at the inverter input terminal from meters at the inverter. We also calculate photocurrent ($\zphi{}$) from processed solar irradiance $S_{I,z}$ and ambient temperature $T_{amb,z}$ data from weather station, making it a \textit{pseudo measurement} of $\IPH{}$ \cite{gilman2015sam}, \cite{masters2013renewable}. 
Note that all measurements represent the aggregate contribution of panels that feed a single inverter.
We model the measurement noise for DC terminal current, DC terminal voltage, and photocurrent with the following random variables:
$\npvv{}, \npvi{}, \nphi{}$, respectively, and we assume Gaussian distribution for all.
%$I_\text{PH}, I_\text{PV}, V_\text{PV}$. To %indicate they are measurements, we note them as 
%$, \zpvi{}, $, respectively.
%The PV system states include open-circuit voltage across the photocurrent source ($\VOC{}$), and DC terminal voltage at the inverter input ($\VPV{}$).

 %the photocurrent shows a linear relationship to solar irradiance, and not sensitive to temperature \cite{ELACHOUBY2018258}. 
% \TODO{Here we need to show it is reasonable to represent $ I_{PH}$ as a pseudo measurement with Gaussian noise}

To include measurements in the circuits framework, we use the substitution theorem to replace any measured component in the circuit with its corresponding measurements. 
Based on the PV system circuit in Fig. \ref{fig:PV_ECF} this yields a new equivalent circuit, as shown in Fig. \ref{fig:PV_MEA_ckt}, which incorporates these measurements.

To represent the physics of the measurement-laden PV system circuit, we include the following KCL-based measurement functions:
\begin{align}
    &-\zphi{}-\nphi{} + \ID{} + \ISH{} + \IPV{} = 0 \label{PV_KCL_1} \\ 
    &-\IPV{} + \zpvi{} + \npvi{} = 0
\end{align}
Now we represent all currents as functions of node voltages and noise variables. 
The current equations for shunt and series resistors are functions of voltage states:
\begin{align}
    &\ISH{} = \frac{\VSH{}}{\RSH{}}\\
    &\IPV{} = \frac{\VSH{} - \VPV{}}{\RS{}}
\end{align}

For diode current, we set $\text{nK}T \rightarrow  \text{a}$ in \eqref{eq:shockley_2} to simplify \eqref{eq:shockley} as they are all constants:
\begin{equation}
    \ID{} = \Izero{}[e^{(\VSH{}/\text{a})} - 1] \label{eq:shockley_2}
\end{equation}
\noindent We use KVL to include the measurement function for DC voltage measurement:
\begin{equation}
    -\VPV{} +  \zpvv{} + \npvv{} = 0 \label{PV_KVL}
\end{equation}
In the estimation problem in Section \ref{sec:model_assem}, we include the following relationships in \eqref{PV_KCL_1} - \eqref{PV_KVL} as constraints for each PV system.

% \TODO{[I think we need to keep a claim here such that the formulation can be further refined, for example, if I have a huge solar farm that is made of 2 different technology and should have a significantly different parameter. It will be awkward to describe them with one general model and intuitively it does not make sense.]}

% \noindent Power balance between the grid and the solar PV models the coupling between the grid and the solar system:
% \begin{equation}
%     P^{\text{AC}} = P^{\text{PV}}
% \end{equation}
% assuming no reactive power is produced and no loss during conversion.

\subsubsection{Battery Measurement Circuits}
\begin{figure}
    \centering
    \includegraphics[scale = 0.09]{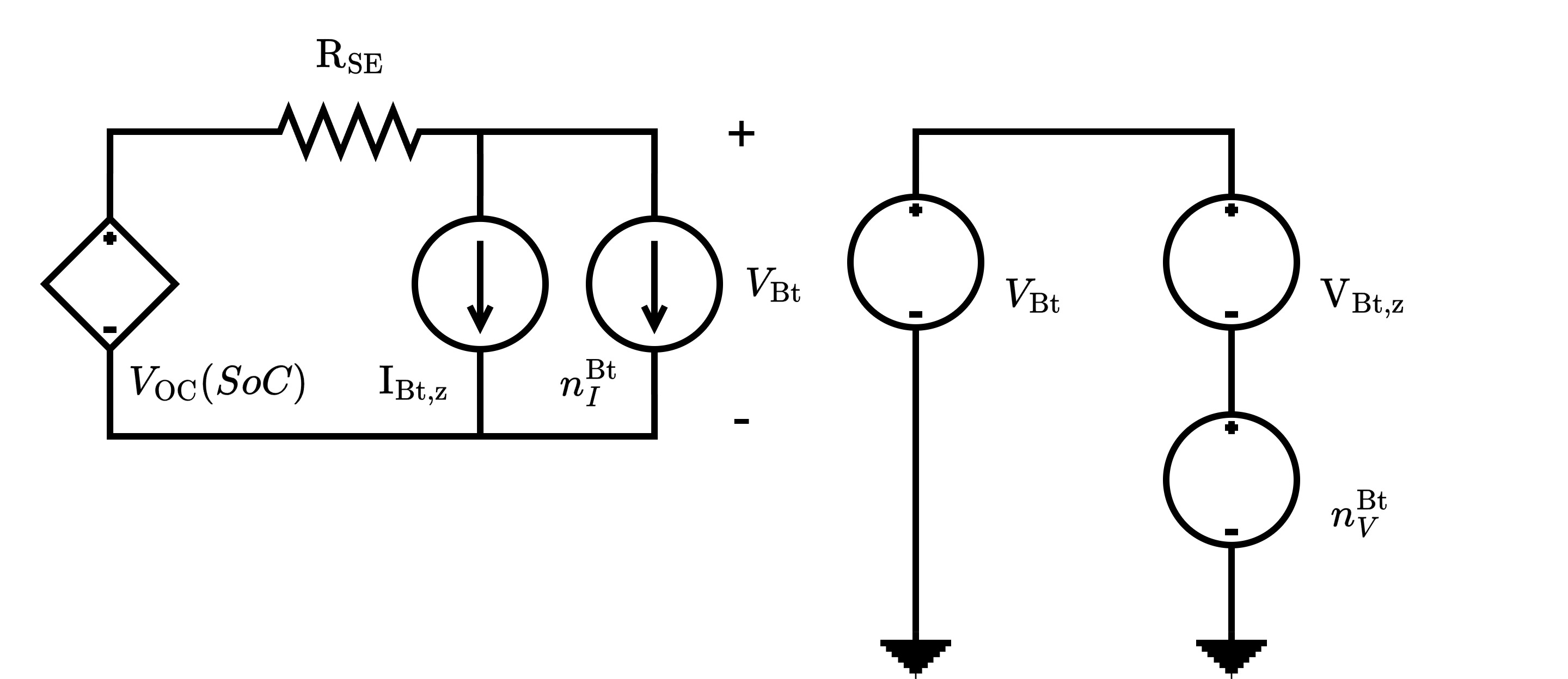}
    \caption{Equivalent circuit of the battery system including measurements.}
    \label{fig:Batt_MEA_ckt}
\end{figure}

%In this work we consider the charging and discharging process of the battery to be a constant power process, that is, power ($\text{P}_{\text{Batt}}$) is a constant during each time step.

\noindent For battery, we consider the following measurements: DC current ($ \zbatti{}$) and DC voltage ($\zbattv{}$) at the inverter input terminal from the meter embedded in the inverter. 
Note that all measurements represent the
aggregate contribution of battery cells and modules that feed a single inverter. 
We model the measurement noise with random variables ($\nbatti{},\nbattv{}$) and assume Gaussian distribution for the noise terms. 
%We estimate the following battery system states: battery state of charge ($\Vsoc{}$), the battery open-circuit voltage ($\VOC{}$), the battery DC terminal voltage at the inverter input ($\VBatt{}$). 

% The variables are unknown states:
% % \TODO{[There is $V_{OC}$ in both PV and Battery, and they were not noted separately, be sure to fix it]}
% \begin{equation*}
%     \Vsoc{},\VBatt{},\VOC{}
% \end{equation*}
% and noise terms as random variables:
% \begin{equation*}
%     \nbatti{}, \nbattv{} 
% \end{equation*}
% The parameters are:
% \begin{equation*}\RSD{},\Ccap{},\text{P}_\text{Bt},\RS{}
% \end{equation*}
Like PV systems, we use the substitution theorem to replace the measured components in the battery circuit with corresponding measurement-based sub-circuits.
With these changes, Fig.\ref{fig:batt_ECF} yields a measurement-based equivalent circuit, as shown in Fig.\ref{fig:Batt_MEA_ckt}.
For measurement functions of the battery circuit, we include a KCL with DC battery current measurement and noise.
\begin{equation}
    -\IBatt{} + \zbatti{} + \nbatti{} = 0 \label{eq:battconst_0}
\end{equation}
\begin{equation}
    \frac{\VBatt{}-\VOC{}}{\RS{}} + \IBatt{} = 0 \label{eq:battconst_1}
\end{equation}
The KVL models the measurement function for DC voltage
measurement:
\begin{equation}
    -\VBatt{} + \zbattv{} + \nbattv{} = 0 \label{eq:battconst_2}
\end{equation}

In general, SoC is hard to directly measure when the system is operating \cite{chen2006accurate}.
However, we can estimate SOC with different methods, as introduced in \cite{zhou2021review}.
We estimate $\Vsoc{}$ with a linear fitted function mapping estimated $\VOC{}$ to estimate $\Vsoc{}$  of the battery, and we also apply the Coulomb counting method to robustify the estimate further. The linear function is:
\begin{equation}\label{eq:f(Vsoc)}
    f_{voc}(\Vsoc{}): \VOC{}(t) = a + b\Vsoc{}(t)
\end{equation}
\noindent for any time step t. While the linear function describes the mapping between $\Vsoc{}$ and $\VOC{}$ well in a certain range (SoC within 20\%-80\%), it is unsuitable for low and high SoC scenario modeling as the relationship gets nonlinear and \eqref{eq:f(Vsoc)} should be replaced by a nonlinear fitted equation. 

We also know the relationship between battery current measurement and $\Vsoc{}$, given by Kirchoff's current law (assuming very large $\RSD{})$:
\begin{align}
    \zbatti{} + \nbatti{} + \Ccap{}\frac{d\Vsoc{}}{dt} &= 0 \label{eq:battconst_3}
\end{align} 
To include this relation in the algebraic constraint set, we apply Trapezoidal approximation, substituting $\zbatti{}$ from \eqref{eq:battconst_0}:
\begin{equation}
    \Vsoc{}(t) - \Vsoc{}(t-\Delta t) = - \frac{\Delta t}{2\Ccap{}}(\IBatt{}(t)+\IBatt{}(t-\Delta t))
\end{equation}
Note that we know ($\Vsoc{}(t-\Delta t)$, $\VOC{}(t-\Delta t)$ from the prior estimation results or the initial condition.
%We represent the time step as $t_\text{step}$.
%KCL equations summarizing the Fig. \ref{fig:Batt_left_MEA_ckt} 
% (assuming very large $\RSD{}$ then $\frac{\Vsoc{}}{\RSD{}} \approx 0$) is as follows:
%Apply the trapezoidal rule:
Next, we represent $\Vsoc{}$ as a function of $\VOC{}$:
\begin{equation}
    %f^{-1}_{voc}(\VOC{}(t)) - f^{-1}_{voc}(\VOC{}(t-\Delta t)) = \frac{\Delta t}{2\Ccap{}}(\IBatt{}(t)+\IBatt{}(t- \Delta t))
    \frac{\VOC{}(t) - \VOC{}(t-\Delta t)}{b} = 
 - \frac{\Delta t}{2\Ccap{}}(\IBatt{}(t)+\IBatt{}(t- \Delta t))
\end{equation}
Finally, substituting $\IBatt{}$ with \eqref{eq:battconst_1}, we get:
\begin{multline}
    \frac{\VOC{}(t) - \VOC{}(t-\Delta t)}{b}= \\
    - \frac{\Delta t}{2\Ccap{}}(\frac{\VOC{}(t)-\VBatt{}(t)}{\RS{}}-\frac{\VOC{}(t-\Delta t)-\VBatt{}(t-\Delta t)}{\RS{}}) \label{eq:batt_soc}
\end{multline}
Now, we include \eqref{eq:batt_soc} in the battery constraint set to estimate $\Vsoc{}$ accurately.
%\begin{figure}
%    \centering
%    \includegraphics[scale = 0.09]{images/Batt_Left_MEA.jpg}
%    \caption{Equivalent circuit of the battery capacity including measurements.}
%    \label{fig:Batt_left_MEA_ckt}
%\end{figure}
% The opencircuituit voltage is a function for SoC:
% \begin{align}
%     V_{OC} = \alpha+\beta V_{SoC}
% \end{align}
% The opecircuitit voltage is unnecessary, or it can be used to check the battery's condition. If the voltage drifted form this function, and the SoC (estimated) is between 30% to 70% then some thing is not correct
% Then with the known power, the relationship between power, voltage, and current can be shown as follows:
% \begin{equation}
%     P_{Batt} = V_{Batt}I_{Batt}
% \end{equation}
% For each component (grid, solar PV system, and battery) we now have a given set of noisy measurements, 
% The power balance between the grid and the battery can be represented as:
% \begin{equation}
%     P_{\text{AC}} =P_{\text{Batt}}
% \end{equation}
% here we assume no reactive power is produced and no loss converts between DC to AC.
% The joint estimation of parameter states calculates the grid states for all buses, as well as for all solar PVs and batteries. The framework can also provide accurate state and parameter estimates when some of the system parameters are unknown.
% \TODO{Make an equivalent circuit graph to include all the information like the NREL inverter modeling paper}
\subsection{Model Assembly} \label{sec:model_assem}
\noindent 

\noindent In this section, we interconnect and aggregate the equivalent circuits in Sections \ref{sec:ECM} and \ref{sec:meas_ckt} to define the estimation problem's constraint set.
The interconnected circuit includes both measurement circuits and those without measurement devices (e.g., zero-injection nodes).

%We have derived the circuit models necessary for formulating the constraint set of the estimation problem defined, including the bulk power grid and grid-tied battery and PV systems, we can now transition to developing the estimation methodology.
%We have also described integrating measurements from RTUs, grid-tied battery systems, and solar PV systems into the equivalent circuit estimation framework.
%In circuit-based estimation, we aggregate the circuit components with measurements on them (i.e., measurement circuits) and other circuits for components without measurements.
\begin{remark}
\textit{To implement the constraint set of the estimation problem, we iterate through the equivalent circuit of each component and add Kirchoff-based network laws into the constraint set. The measurement information is embedded within the equivalent circuits.}
\end{remark}

\begin{figure*}[ht]
  \includegraphics[scale = 0.05]
  {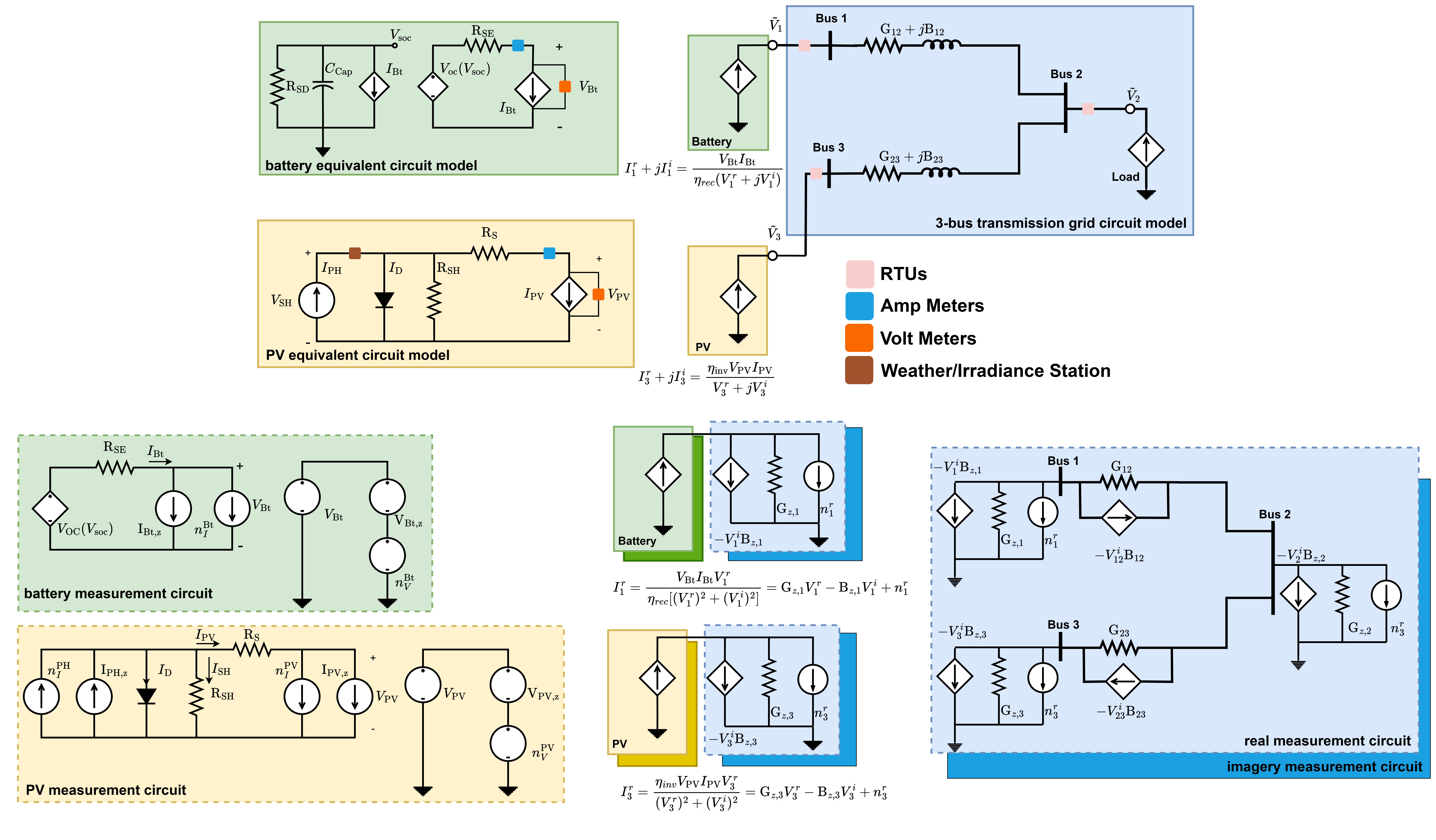}
  \centering
  \caption{This figure shows the combined 3-bus circuit example when the battery charges from the grid and the PV discharges to the grid. The upper half of the graph is ECM while the lower half is the measurement circuit.}\label{fig:Full_combined_circuit}
\end{figure*}

As an example, consider the 3-bus example in Fig. \ref{fig:Full_combined_circuit}. 
An inverter-connected battery system (green box) is connected to Bus 1 of the grid model (blue box), and an inverter-connected solar PV facility (yellow box) is connected to Bus 3 of the grid model. A load is connected to the Bus 2 of the grid.

%where the power output on node 1 is from a battery system, and the node is from an inverter-connected PV system.
%To interconnect individual circuits and build the constraint set, ...
%In the context of this paper, we consider different sets of circuits:
%In the following paragraph, we will first explain how the circuits connect to each other to form a full equivalent circuit of the whole system,  then develop the Ckt-SE problem from the circuits, and finally, we extend Ckt-SE to the Ckt-PSE problem when erroneous parameters exist in the formulation. We solve both problems as non-convex constrained optimizations.
%Now we included three groups of subcircuits represented as ECF:
%\begin{itemize}
%    \item The grid equivalent circuit
%    \item The solar PV equivalent circuit
%    \item The battery equivalent circuit
%\end{itemize}
%An example of a simple 3-bus case with all three circuits is Fig.\ref{fig:Full_combined_circuit}. The system consists of three main components: a battery storage system (green box), a solar PV system (yellow box), and the transmission grid they are connected to (blue box). 

%Note that various sensors instrument all three systems, and their coupling is modeled using controlled sources.

To run an estimation study, we replace the measured components in Fig. \ref{fig:Full_combined_circuit} [top] with measurement-based equivalent circuits from Section \ref{sec:meas_ckt} (see Fig. \ref{fig:Full_combined_circuit} [bottom]). 
We connect them together as shown in the dashed boxes in Fig. \ref{fig:Full_combined_circuit}.
We then write the IV relationships for each component in the aggregated circuit to describe the measurement functions or the constraint set for the estimation problem.
%Each of these components (represented by solid-line boxes) contains a corresponding measurement circuit (represented by dashed-line boxes). The parts in the yellow and green box connected to the grid circuit are equivalent current source account for solar PV and battery system power injection behavior to the grid. 
% This figure shows the case that the battery is charging from the grid and the PV is discharging to the grid. 
% They are power source coupled to their own component circuit and represented as negative load, thus the resistors in these circuits can have negative values when PV generating or battery discharging. 
% \begin{figure*}[h]
%   \includegraphics[width=20cm,height=13cm]{images/Combined_circuit_1016.jpg}
%   \caption{Combined 3-bus circuit example}\label{fig:Full_combined_circuit}
% \end{figure*}

%All components in each circuit should satisfy KCL and KVL equations. When we wrap these physics rules as optimization problems in the next section, they act as the constraints.

\subsection{Circuit-based State Estimation as an Optimization} \label{sec:estimation_routines}

\noindent \textbf{Forming optimization problem:}
Next, we formulate the optimization routines for the estimation problem. 
We formulate two sets of optimizations: i) \textit{stand-alone} grid, battery, and PV system estimation routines, where each system independently models and estimates its states and ii) \textit{combined} optimization routine, where grid, PV, and battery systems are modeled and estimated together.
%When forming the circuit-theoretic parameter state joint estimation with PV and battery measurements optimization problem, we categorize the methods as stand-alone formulation and combined formulation based on whether the formulation couples to the grid. Separated formulation solves each one of the components in parallel, thus we also name it \text{stand-alone method} as each component estimates its own state with its own measurements. While combined formulation solves the whole system as a large circuit, all components are coupled to each other thus we also call it \text{grid co-estimation (combined)} method.\\

\noindent \textit{1. Stand-alone Estimation Routines}

\noindent \textbf{Stand-alone grid formulation:}
The stand-alone grid estimation is introduced in Section \ref{sec:grid ECF}. The optimization minimizes noise realization in RTU measurements subject to grid network constraints:
\begin{subequations} \label{opt:grid_sep}
\begin{equation}
       \cktse{}:\sum_{m \in{\mathcal{M}}}f_{m}^{\text{Grid}}(n)
\end{equation}
subject to:
\begin{equation} 
    h_{{i}}^{\text{Grid}}(V,n) = 0  \;\; \forall i \in {\mathcal{N}\backslash {\mathcal{N}_{ZI}}} \label{eq:Std_kcl}\\
\end{equation}
\begin{equation}
    h_{i}^{\text{Grid}}(V) = 0 \;\; \forall i \in {\mathcal{N}_{ZI}} \label{eq:Std_zi} \\
\end{equation}
\end{subequations}
See \cite{jovicic2020enhanced},\cite{CircuitTheoretic2020},  and \cite{sang2024circuit} to learn construction of objective function $f^\text{Grid}(n)$ and constraint set \eqref{eq:Std_zi} and \eqref{eq:Std_kcl} for injection and zero injection nodes.\\
\noindent \textbf{Stand-alone PV formulation:}
For stand-alone PV estimation, we minimize the weighted-noise realizations from PV system measurements: output DC current$\zpvi{}$ and voltage ($\zpvv{}$), and photocurrent ($\zphi{}$).
% Just like the grid SE problem covered by \cite{CircuitTheoretic2020} and introduced in section \ref{sec:grid ECF}, we can also formulate the HES SE in the circuit formulation. 
The problem formulation is as follows: 
\begin{subequations} \label{opt:PV_sep}
\begin{equation}
       \pvse{}: \min_{V,n} f^\text{PV}(n)
\end{equation}
subject to:
\begin{equation}\label{eq:cktSE_hrtu}
         h^\text{PV}(V,n) = 0\\
\end{equation} the constraint \eqref{eq:cktSE_hrtu} includes PV system's circuit's IV-relationships, described in (\ref{PV_KCL_1}) - (\ref{PV_KVL}) and the objective has the following form:
\begin{multline}\label{PV_obj}
        f^\text{PV}(n) = \min_{V,n}\left( w^\text{PV}_\text{I}{(\npvi{})}^2 +{w^\text{PV}_\text{V}(\npvv{})}^2 \right.\\ 
        \left. +  w^\text{PH}_\text{I}{(\nphi{})}^2 \right)  
\end{multline}
\end{subequations}
In (\ref{PV_obj}), the weight ($w$) accounts for the spread in measurement accuracy and for measurement $i$, $w_i$ is defined by the inverse of the measurement distribution variance ${\sigma _i}^{-2}$.

\noindent \textbf{Stand-alone battery formulation:}
For stand-alone battery system estimation, we estimate the battery states, including SoC, given DC output voltage ($\zbattv{}$) and current ($\zbatti{}$) measurements:
\begin{subequations} \label{opt:Batt_sep}
\begin{equation}
       \btse{}:\min_{V,n} f^\text{Bt}(n)
\end{equation}
subject to:
\begin{equation}\label{eq:constBatt_matrix}
         h^\text{Bt}(V,n) = 0\\
\end{equation}
\eqref{eq:constBatt_matrix} includes constraints from (\ref{eq:battconst_1}) to (\ref{eq:battconst_3}) that describe the measurement functions for the battery.
The objective is given in \eqref{Batt_obj}, minimizing the weighted noise realizations in the battery measurements:
\begin{equation}
    f^\text{Bt}(n) = \min_{V,n} \left(w^\text{Bt}_\text{I}{(\nbatti{})}^2 +{w^\text{Bt}_\text{V}(\nbattv{})}^2 \right) \label{Batt_obj}\\
\end{equation}
\end{subequations}

% We can note this optimization problem as $\mathbf{{P}_{PV,i}}$ for the $\mathbf{i^{th}}$ solar system.
% Similarly, for each battery storage system, we can also construct an optimization problem $\mathbf{{P}_{Batt,j}}$ for the $\mathbf{j^{th}}$ battery system.

\noindent Overall, the stand-alone routines are easier to scale up as states for each system are estimated independently. However, the results of each subcircuit are not guaranteed to be consistent at the boundaries, and the impact of bad-data is significant in the estimate quality. Therefore, we formulate the estimation routine for combined PV, battery, and grid systems to address these challenges.

\noindent \textit{2. Combined Estimation Framework}: $\cktser{}$

%\noindent\textbf{Combined formulation ($\cktser{}$):}
\noindent Instead of estimating the states of solar PV or battery subsystems independently from the grid, in the combined formulation, we couple the utility-scale battery and PV systems with the bulk grid model, and we perform estimation on the coupled model.
We term this formulation: $\cktser{}$. 
To couple the systems, we aggregate the equivalent circuits for PV, battery, and grid components in Section \ref{sec:meas_ckt} and introduce a new set of coupling constraints that reconcile the physics at the point of interconnection (POI).
The coupling constraints enforce the power balance between the solar PV and battery and grid subsystems.
% \TODO{We describe the whole solar farm and whole battery measurements with a lumped model each, the question is, where do I get this measurement, is it just an average of the panels? The parameters are also considered to be using a spec sheet data, we assume that the parameters are correct, and we assume the model describes the system well.}
%By combining solar PV, battery, and grid, a new set of constraints becomes available. For each solar PV or battery connected to the grid, there is a set of power injection constraints coupling them to the grid.
Inverters couple the various subsystems, and the behavior is dependent on the choice of control settings.
For solar PV subsystems, we assume MPPT output. 
For battery subsystems, we assume constant PQ discharge or charge cycle based on an external dispatch schedule.
For this work, we also assume unity power factor operation of the inverters and therefore reactive power transfer between the subsystems is assumed zero.

The grid circuit has both real and imaginary components.
The AC power is obtained from complex voltage and current product.
Subsequently, for solar PV at bus $b$, and MPPT control at unity power factor:
\begin{equation}
    h^{\text{inv}}: \left\{
    \begin{aligned}
    &\eta_{inv}\IPV{}\VPV{}-(V_{b}^{r}I_{b}^{r} + V_{b}^{i}I_{b}^{i}) = 0\\
    &V_{b}^{i}I_{b}^{r} - V_{b}^{r}I_{b}^{i} = 0
    \end{aligned}
    \right.
    \label{eq:PV_inv}
\end{equation}    
For battery system at bus $d$, the constant power relationship is given by:
% \begin{equation}
%      \eta_{inv}^D\IBatt{}\VBatt{} -(V_{d}^{r}I_{d}^{r} + V_{d}^{i}I_{d}^{i}) = 0 \label{eq:battery_discharge_P}
% \end{equation}
% \begin{equation}
%      \IBatt{}\VBatt{} -\eta_{inv}^C(V_{d}^{r}I_{d}^{r} + V_{d}^{i}I_{d}^{i}) = 0\label{eq:battery_charge_P}
% \end{equation}
% \begin{equation}
%     V_{d}^{i}I_{d}^{r} - V_{d}^{r}I_{d}^{i} = 0\label{eq:battery_charge_Q}
% \end{equation}

\begin{equation}
    h^{\text{inv}}:\left\{
    \begin{aligned}
    &\eta_{inv}\IBatt{}\VBatt{} -(V_{d}^{r}I_{d}^{r} + V_{d}^{i}I_{d}^{i}) = 0, & \textrm{\textbf{if} discharging}\\
     &\IBatt{}\VBatt{} -\eta_{rec}(V_{d}^{r}I_{d}^{r} + V_{d}^{i}I_{d}^{i}) = 0, & \textrm{\textbf{if} charging} \label{eq:battery_inv_charging}\\
    &V_{d}^{i}I_{d}^{r} - V_{d}^{r}I_{d}^{i} = 0\\
    \end{aligned}
    \right.
\end{equation}

In the combined setup, these inverter equations couple the stand-alone voltage and current measurements with  RTU measurements recorded at the sub-station (see \eqref{eq:feature_trans}, \eqref{eq:rtu_injection_1} and \eqref{eq:rtu_injection_2}).
Note that both rectification (charge) and inversion (discharging) efficiencies are functions of operating power and are discussed in the inverter section \ref{sec:inverter_modeling}. 
As solar PV does not consume power from the grid, we only consider injecting power balance \eqref{eq:PV_inv}. 
The coupling constraints for battery \eqref{eq:battery_inv_charging} considers both charging and discharging scenarios. For example, Fig. \ref{fig:Full_combined_circuit} shows the battery charging with \eqref{eq:battery_inv_charging} and PV discharging with \eqref{eq:PV_inv} at bus 1 and bus 3, respectively.

%As the current terms are functions of bus voltages as shown in equation \eqref{eq:rtu_injection_1} and \eqref{eq:rtu_injection_2}, the variables in the power mismatch equations are only voltages and noise random variables. We consider the power balance equations \eqref{eq:PV_inv} to \eqref{eq:battery_inv_charging} to be included in the corresponding component constraint set.
% Thus we represent the collection of these power balances as follows:
% \begin{equation}
%     h_\text{power}(V,n) = 0\\
% \end{equation}

We summarize the combined state estimation of PV, battery and grid systems in \eqref{opt:reckt_se}:
\begin{subequations} \label{opt:reckt_se}
\begin{equation}
        \cktser{}:\min_{V,n} f(n)
\end{equation}
subject to:
\begin{equation} 
    h_{{i}}^{\text{Grid}}(V,n) = 0  \;\; \forall i \in {\mathcal{N}\backslash {\mathcal{N}_{ZI}}} \label{eq:grid_kcl}\\
\end{equation}
\begin{equation}
    h_{i}^{\text{Grid}}(V) = 0 \;\; \forall i \in {\mathcal{N}_{ZI}} \label{eq:grid_zi} \\
\end{equation}
\begin{equation}
         h_{k}^{\text{PV}}(V,n) = 0 \;\; \forall k \in {\mathcal{K}} \label{eq:pv_eq}\\
    % \end{split}
\end{equation}
\begin{equation}
    h_{l}^{\text{Bt}}(V,n) = 0  \;\; \forall l \in {\mathcal{L}}
    \label{eq:bt_eq}\\
\end{equation}
\begin{equation}
    h_{i}^{\text{inv}}(V) = 0 \;\; \forall i \in {\mathcal{K}} \cup {\mathcal{L}} \label{eq:grid_inv} \\
\end{equation}
% \begin{align}
%     h_k,\text{power}(V,n) &= 0 \;\; \forall k \in \mathcal{K}\\\
%     h_l,\text{power}(V,n) &= 0 \;\; \forall l \in \mathcal{L}
% \end{align}
where, $i$ is a grid bus index and $k$ and $l$ are PV and battery circuit-index. ${\mathcal{K}} $ and ${\mathcal{L}}$ are sets of solar PV and battery systems, respectively. $\mathcal{N}\backslash \mathcal{N}_{ZI}$ is a set of grid buses with RTU on it, and ${\mathcal{N}_{ZI}}$ are unmeasured zero-injection buses. 
The constraints for grid, battery, and solar PV are identical to the stand-alone method \eqref{eq:cktSE_hrtu} and \eqref{eq:constBatt_matrix}.
Except, in the combined approach, we include the extra inverter-based coupling constraints for PV and battery systems. %The real and reactive power constraints for PV  are \eqref{eq:PV_inv}. For charging and discharging a battery the constraints are \eqref{eq:battery_inv_charging}. For each solar PV and battery in the system it provides a set of constraints as shown in the separated formulation, and all components contribute the constraints as shown in this formulation. 
%For brevity, we term the set of all constraints in $\cktser{}$ except \eqref{eq:grid_zi} $h^\text{Re}(V,n)$.

The norm-2 minimization in objective has the following form:
\begin{multline}
    f(n) = \min_{V,n} \left( \sum_{k \in  {\mathcal{K}}}f_{k}^{\text{PV}}(n) + \sum_{l \in  {\mathcal{L}}}f_{l}^{\text{Bt}}(n)\right.\\
    \left. + \sum_{m \in{\mathcal{M}}}f_{m}^{\text{Grid}}(n) \right)
\end{multline}
\end{subequations}
\noindent \textit{3. Include erroneous parameters}: $\cktpser{}$\\
In the combined $\cktser{}$ and stand-alone algorithms, the circuit parameters (${\mathcal{P}}$) like $\RS{},\RSH{}$ for solar PV circuits and $\RSE{}$ for battery circuits, are considered known. 
However, in many real-world instances, these parameters might be erroneous. 
For solar PV systems, $\RS{}$ can deviate from the fact sheet number due to the adjustments on wiring, replacement or aging of cable, and other reasons such as human error during data input. 
For battery systems, the internal resistance changes with the battery system's age and the cell's temperature.
% When there are erroneous parameters in the system, one can pause the SE and stop getting estimates until the correct parameters are regained. There are also cases, like system monitoring, in which we still want estimates from the system with some level of accuracy although it will be less reliable.
To address this, we modify our $\cktser{}$ to include erroneous parameters.
In the algorithm, we replace the erroneous parameters (${\mathcal{P}}_\text{U}$) as unknown variables and solve them within the optimization problem. 
%When a circuit has erroneous parameters (${\mathcal{P}} \in {\mathcal{P}}_\text{U}$), the parameter variables replace the erroneous parameters and then get estimated, we name this algorithm $\cktpse{}$ to differentiate from cases without erroneous parameters. 
We term the algorithm with this adjustment as $\cktpser{}$ to differentiate from $\cktser{}$ algorithm without erroneous parameters. 
We summarize the $\cktpser{}$ algorithm in Algorithm \ref{alg:Re-CktPSE}.
\begin{algorithm}[ht]
\caption{\textbf{$\cktpser{}$}}\label{alg:Re-CktPSE}
\begin{algorithmic}[1]
\State \textbf{Input:} ${\mathcal{P}},{\mathcal{P}}_\text{U},{\mathcal{Z}}, \boldsymbol{\mathcal{W}},{\mathcal{K}},{\mathcal{L}},{\mathcal{M}},{\mathcal{G}}$
% \State $(\zbatti{},\zbattv{})_k \gets \boldsymbol{\mathcal{Z}} \;\;\; \forall k \in \mathcal{K}$
% \State $(\zpvi{},\zpvv{},\zphi{})_l \gets \boldsymbol{\mathcal{Z}} \;\;\; \forall l \in \mathcal{L}$
% \State $(\Gkz{},\Bkz{})_m \gets \boldsymbol{\mathcal{Z}} \;\;\; \forall m \in \mathcal{M}$
% \State $\mathcal{P} \gets \boldsymbol{\mathcal{P}}$
\State \textbf{Set:} ${\mathcal{H}}=\phi$ (empty constraint set), $f(n)=0$ (obj. fcn)
\vspace{4pt}
\hrule
\vspace{4pt}
\State \textbf{Step 1:} Iterate through components with meas (${\mathcal{K}} \cup {\mathcal{L}} \cup {\mathcal{M}}$) and add terms to the constraint set ${\mathcal{H}}$:
\For{$ m \in {\mathcal{K}} \cup {\mathcal{L}} \cup {\mathcal{M}}$} \Comment{for every measurement ckt.}
\State $\mathcal{Z}_m \gets {\mathcal{Z}}$\Comment{get ckt. meas from meas set}
\State $\mathcal{P}_m \gets {\mathcal{P}}$\Comment{get ckt params from param set}
\State \textit{Add measurement func. to the constraint set:}
\If{$\exists \mathcal{P}_m \in {\mathcal{P}}_\text{U}$}\Comment{If parameter unknown} 
    \State ${\mathcal{H}} \leftarrow {\mathcal{H}} \cup h_m(V,n,\mathcal{P}_m) $ 
    % \Comment{with unknown parameters}
\Else{}
    \State ${\mathcal{H}} \leftarrow {\mathcal{H}} \cup h_m(V,n) $ 
    % \Comment{without unknown parameters}
\EndIf
\State \textit{Form component objective:}
\State $w_m \gets \boldsymbol{\mathcal{W}}$ \Comment{get weights from covariance matrix}
\State $f(n) += w.n^2 \quad \forall w \in w_m, n \in n_m$
%\State $n \gets n_m$ \Comment{add noise term to noise vector}
\EndFor
\vspace{4pt}
\hrule
\vspace{4pt}
\State \textbf{Step 2:} Iterate through components without measurements $({\mathcal{G}})$ and add terms to the constraint set ${\mathcal{H}}$:
\For{$j \in {\mathcal{G}}$} \Comment{for non-measured grid components}
\State $\mathcal{P}_j \gets {\mathcal{P}}$ \Comment{get params form param set}
\If {$\exists \mathcal{P}_j \in {\mathcal{P}}_\text{U}$} 
    \State ${\mathcal{H}} = {\mathcal{H}} \cup h_j(V,\mathcal{P}_j) $ 
    % \Comment{with unknown parameters}
\Else{}
    \State ${\mathcal{H}} = {\mathcal{H}} \cup h_j(V) $ 
    % \Comment{without unknown parameters}
\EndIf
\EndFor
\State constraint set ${\mathcal{H}}$ is complete with all terms in \ref{alg:Re-CktPSE}
\vspace{4pt}
\hrule
\vspace{4pt}
\State \textbf{Step 3:} Solve the optimization to obtain estimates
\State min $f(n)$
\Statex s.t. ${\mathcal{H}}(V, n, \mathcal{P}_\text{U}) = 0$
\State \textbf{Output:} $\hat{V}, \hat{\mathcal{P}}_\text{U}$
\end{algorithmic}
\end{algorithm}

\section{Experiment Design}
% \TODO{Cover the measurement accuracy here, IEEE - 14 bus with HES on some buses}

% \TODO{For now the code works (produce meaningful estimation) for at most 118 case, shows index error for 2869 buses}
\noindent To study the performance of the proposed approaches, we consider several experiments, categorized by 2 setups, which operate on 3 scenarios. The setups are described below:

\begin{itemize}
    \item \textbf{Setup 1 (S1):} Stand-alone estimation of battery (\btse{}), PV (\pvse{}) and bulk grid system (\cktse{}) states
    \item \textbf{Setup 2 (S2):} Combined estimation of battery, PV, and bulk grid system states ($\cktser{}$ and $\cktpser{}$)
\end{itemize}
\textbf{S1} estimates each component independently following methodology in Section \ref{sec:estimation_routines}1. \textbf{S2} performs combined estimation based on methodology in Section \ref{sec:estimation_routines}2.
\setlength{\tabcolsep}{6pt}
\begin{table}[ht]
    \caption{Test Cases Used}
    \centering
    \renewcommand{\arraystretch}{1.1} % Adjust the vertical spacing between rows
    \begin{adjustbox}{max width=\textwidth}
    \begin{tabular}{@{}lcccc@{}}
        \hline
        \makecell{\textbf{Test Case}} & \makecell{\textbf{Network}} & \makecell{\textbf{Number of} \\ \textbf{PV}} & \makecell{\textbf{Number of} \\ \textbf{Battery}} & \makecell{\textbf{Unknown} \\ \textbf{Parameter}} \\
        \hline
        \case{1}{} & IEEE-118 & $11$ & $0$& No\\
        \case{2}{a} & 2869pegase & $10$ & $2$& No\\
        \case{2}{b} & 2869pegase & $10$ & $1$& Yes\\
        \case{3}{} & ACTIVSg10k & $100$ &$0$ & No\\
        \hline
    \end{tabular}
    \end{adjustbox}
    \label{tab:Case_setup}
\end{table}
\noindent We consider three scenarios to compare the performance of stand-alone setup \textbf{S1} against the combined setup \textbf{S2}:
\begin{itemize}
    \item In scenario A, we estimate the system states when all parameters are known, and the measurements only include white noise
    \item In scenario B, we estimate the system states when all parameters are known, and certain measurements include biased bad data
    \item In scenario C, we estimate the system states with unbiased noisy measurements but study the estimation robustness against \textit{certain} erroneous parameters 
\end{itemize}

\noindent We use 4 test cases to study the three scenarios. The cases are described in Table. \ref{tab:Case_setup}. All test cases are in per unit system with a base of 100 MW. Cases 1 through 4 study scenario A. Case 2 focuses on scenario B, and Case 3 focuses on scenario C. In all experiments, the estimation quality of the grid parameters and states is not the key focus. See prior work in that regard \cite{sang2024circuit}.

%\noindent Further, we show the consistency of estimates at the interconnection (e.g., power output estimate PV is equivalent to power injection on the grid node) with the combined setup.

%\noindent \textbf{Experiment setup:}
%We designed the experiments to compare three main scenarios. 
%The three scenarios are test cases with no bad data, test cases with bad data, and test cases with no bad data but with unknown parameters. 

For each scenario, we ran 100 instances ($N_s$) with random noise realizations to produce statistically relevant results for estimation with the two setups.
We assume all injection nodes are measured on the grid with an RTU, and all solar PV or battery subsystems have measurements available. 
The choice of measurements is described in Section \ref{sec:meas_ckt}.
To generate synthetic measurements, we first simulate the stand-alone and combined systems and record the simulation solution.
Next, we draw noise samples from a zero-mean Gaussian distribution with a standard deviation
value of 0.001 \cite{CircuitTheoretic2020} for RTU measurements and a deviation value of 0.1 for inverter readings and add those to the simulation results.
%Bad data injection is done by adding noise that is from a high variance distribution.
For scenarios including bad data, we assume that bad data comes from a PV or battery system meter. The bad measurement is assumed to be biased in magnitude and output non-zero mean noise. 

Scenario C studies the estimation performance in the presence of erroneous parameters. 
We induce errors in $\RS{}$ parameter in the solar PV system. 
% and $\RSE{}$ parameter in the battery system.
We choose these as erroneous as they are more likely to be incorrect than other parameters: $\RS{}$ account for interconnection loss, which varies from site to site. 
%$\RSE{}$ accounts for internal loss, which depends on the battery temperature and is also an indicator of battery life. 
%In this work, we assume constant resistance in both cases considering the experiment should represent a short steady state time period of a system.

\noindent \textbf{Result metrics:}
We compare the experiment outcome based on the following error metrics: normalized root mean square error (NRMSE), variance ($\sigma^2$), and estimation error (\%). The estimation error is how much the estimation differs in percentage compared to the true value; the formula is shown in  \eqref{eq:est_error}, where $\hat{y} \; \text{and} \; y_{true}$ represents the estimated value and the true value of the unknown state. 
%\TODO{[Note, declare here that our solar panels and battery storage systems are assumed to be the same criteria but parameter in size due to the limitation of data in hand]}
% \begin{FlushLeft}

\begin{align}
    %\text{Estimation Error} &= \frac{\sum^{n}_{i=0} |\textbf{y}_i - {\boldsymbol{\mu}}|}{n}
    \text{Estimation Error}(\%) &= \frac{(\hat{y} - {y_{true})}}{y_{true}} 100 \label{eq:est_error}
\end{align}

\begin{equation}
\begin{split}
       RMSE_{\mathbf{y}} = \sqrt{\frac{\sum^{N_s}_{n=1} \sum^{{{N_c}}}_{c=1} (\hat{y}_{nc} - {y}_{true,c})^2}{N_s N_c}}
\end{split}
\end{equation}
\begin{equation}
\begin{split}
    NRMSE_{\mathbf{y}} = \frac{RMSE_{\mathbf{y}}}{\hat{y}_{avg}}\label{eq:NRMSE}
\end{split}
\end{equation}
\begin{equation}
\begin{split}
    \sigma^2_{c} = {\frac{\sum^{{N_s}}_{n=0} (\hat{y}_{nc} -\hat{y}_{avg,c})^2}{N_s}} 
\end{split}
\end{equation}
\begin{equation}
\begin{split}
    \sigma^2_{avg} = \frac{\sum^{{{N_c}}}_{c=1}\sigma^2_{c}} {{N_c}} 
\end{split}
\end{equation}
% \end{FlushLeft}
Here $\hat{y}_{nc}$ is the estimate of $n^{th}$ run of component $c$, which can be PV, battery, or grid node voltage.
$\hat{y}_{avg,c}$ is the average value of estimates across the $N_s$ runs of simulations of component $c$. The NRMSE and variance for grid nodes, battery, or solar PV state and parameter estimates are all calculated in the same manner.

% The power mismatch shows how consistent the estimation of the components is to the grid estimation, the lower the more consistent. For example, the power mismatch at bus $d$ with a solar PV connected to the grid can be calculated as shown in \eqref{eq:pow_miss}.
% \begin{equation}\label{eq:pow_miss}
%     P_{d,\text{diff}} = |\hat{V}_\text{PV}\frac{(\hat{V}_\text{OC} - \hat{V}_\text{PV})}{\text{R}_\text{S}} -(\hat{V}^{r}_{d}I^{r}_{d} + \hat{V}^i_{d}I^i_{d})|
% \end{equation}
% \begin{equation}\label{eq:pow_error}
%     P_{\text{error}} = |\hat{V}_\text{PV}\frac{(\hat{V}_\text{OC} - \hat{V}_\text{PV})}{\text{R}_\text{S}} - P_{\text{true}}|
% \end{equation}
% Where $\hat{V}_\text{PV},\hat{V}_\text{OC},\hat{V}^{r}_{d},\hat{V}^i_{d}$ are the estimates of state variables $\VPV{},\VOC{},V_d^{r},V_d^{i}$ for solar PV and the grid. The current $I^d_{i}, I^d_{r}$ are real and imaginary net current injections of bus $d$, a function of state variables of bus $d$ and buses connected to bus $d$. 
% To compare across different cases we calculate the power mismatch ratio (PMR) for bus $d$ as shown in \eqref{eq:pow_miss_ratio}.
% \begin{equation}\label{eq:pow_miss_ratio}
%     PMR(\%) =\frac{P_{d,\text{diff}}}{\hat{V}^{r}_{d}I^{r}_{d} + \hat{V}^i_{d}I^i_{d}}  
% \end{equation}
%  We use $\text{PMR}_{ave}$ to represent the average PMR in multiple tests. $PMR$ shows the inconsistency between PV and grid estimation, while the $P_{d,\text{error}}$ shows the power estimation accuracy.
The error can be positive or negative when comparing the battery SoC estimation. Because the error accumulates by time step, we use the absolute error to keep the figure consistent:
\begin{equation}
    \text{Absolute Error (\%)} = |\frac{(\hat{y} - {y_{true})}}{y_{true}} 100| \label{eq:abs_error}
\end{equation}

\section{Experiment Results}
\noindent The following section shows state and parameter estimation results in three main scenarios: for test cases with and without bad data in measurements and test cases with erroneous parameters. 
We compare the performance of stand-alone algorithms against $\cktser{}$ for the first two scenarios and against $\cktpser{}$ for the third scenario. 
We summarize the solving time for various experiments to show that $\cktser{}$ and $\cktpser{}$ scales well.

\subsection{State estimation of solar PV and battery system without bad data} 
\noindent Both stand-alone and combined algorithms can estimate the state of the battery and PV system. 
To compare their performance, we summarize the PV state estimation accuracy for all four test cases in Table \ref{tab:State_Est_accuracy}. 
In \case{1}{}, \case{2}{a}, \case{2}{b}, and \case{3}{}, we observe that both stand-alone and $\cktser{}$ perform well; however, $\cktser{}$ consistently outperforms stand-alone algorithm.
Note that in Table \ref{tab:State_Est_accuracy} through Table \ref{tab:State_Est_accuracy_batt_biased}, DC power refers to the power injection from the PV system to the inverter and also the DC charging power from the rectifier to the battery, defined by $\IPV{}\VPV{}$ for solar PV and $\IBatt{}\VBatt{}$ for the battery.
\begin{table}[ht]
\caption{PV State estimation results (w/o bad data).}\label{tab:State_Est_accuracy} 
\centering
\scriptsize
\renewcommand{\arraystretch}{1.5} % Adjust vertical spacing between rows
\begin{tabular}{@{}l@{\hspace{2pt}}cccccc@{}}
\hline
\multirow{2}{*}{\textbf{Algo./Case}} & \multicolumn{2}{c}{$\VPV{}$} & \multicolumn{2}{c}{$\VSH{}$} & \multicolumn{2}{c}{DC Power} \\ 
\cmidrule(lr){2-3} \cmidrule(lr){4-5} \cmidrule(lr){6-7}
 &  NRMSE& $\sigma^2_\text{avg}$ &  NRMSE & $\sigma^2_\text{avg}$ & NRMSE & $\sigma^2_\text{avg}$ \\ \hline
\pvse{}/\case{1}{} & $2.7\text{E-}02$ & $2.5\text{E-}02$ & $9.2\text{E-}01$ & $8.6\text{E-}01$ & $2.6\text{E-}01$ & $6.7\text{E-}05$ \\
$\cktser{}$/\case{1}{}     & $2.6\text{E-}01$ & $2.4\text{E-}02$ & $8.5\text{E-}01$ & $8.0\text{E-}01$ & $2.7\text{E-}01$ & $6.4\text{E-}05$ \\ \hline
\pvse{}/\case{2}{a}  & $2.7\text{E-}02$ & $2.5\text{E-}02$ & $9.2\text{E-}01$ & $8.6\text{E-}01$ & $2.7\text{E-}02$ & $5.7\text{E-}05$ \\
$\cktser{}$/\case{2}{a}     & $2.3\text{E-}02$ & $2.1\text{E-}02$ & $6.7\text{E-}01$ & $6.3\text{E-}01$ & $2.7\text{E-}02$ & $6.4\text{E-}05$ \\ \hline
\pvse{}/\case{2}{b} & $2.6\text{E-}02$ & $3.3\text{E-}02$ & $8.2\text{E-}01$ & $1.5\text{E+}00$ & $3.9\text{E-}02$ & $1.6\text{E-}04$ \\
$\cktser{}$/\case{2}{b}    & $1.2\text{E-}02$ & $1.3\text{E-}02$ & $1.7\text{E-}01$ & $1.2\text{E-}01$ & $1.9\text{E-}02$ & $2.4\text{E-}05$ \\ \hline
\pvse{}/\case{3}{} & $2.8\text{E-}02$ & $2.5\text{E-}02$ & $9.6\text{E-}01$ & $9.0\text{E-}01$ & $9.6\text{E-}03$ & $6.3\text{E-}05$ \\
$\cktser{}$/\case{3}{}    & $2.5\text{E-}02$ & $2.3\text{E-}02$ & $8.0\text{E-}01$ & $7.6\text{E-}01$ & $9.9\text{E-}03$ & $6.7\text{E-}05$ \\ \hline
\end{tabular}
\end{table}
\begin{figure}
    \centering
    \includegraphics[scale = 0.33]{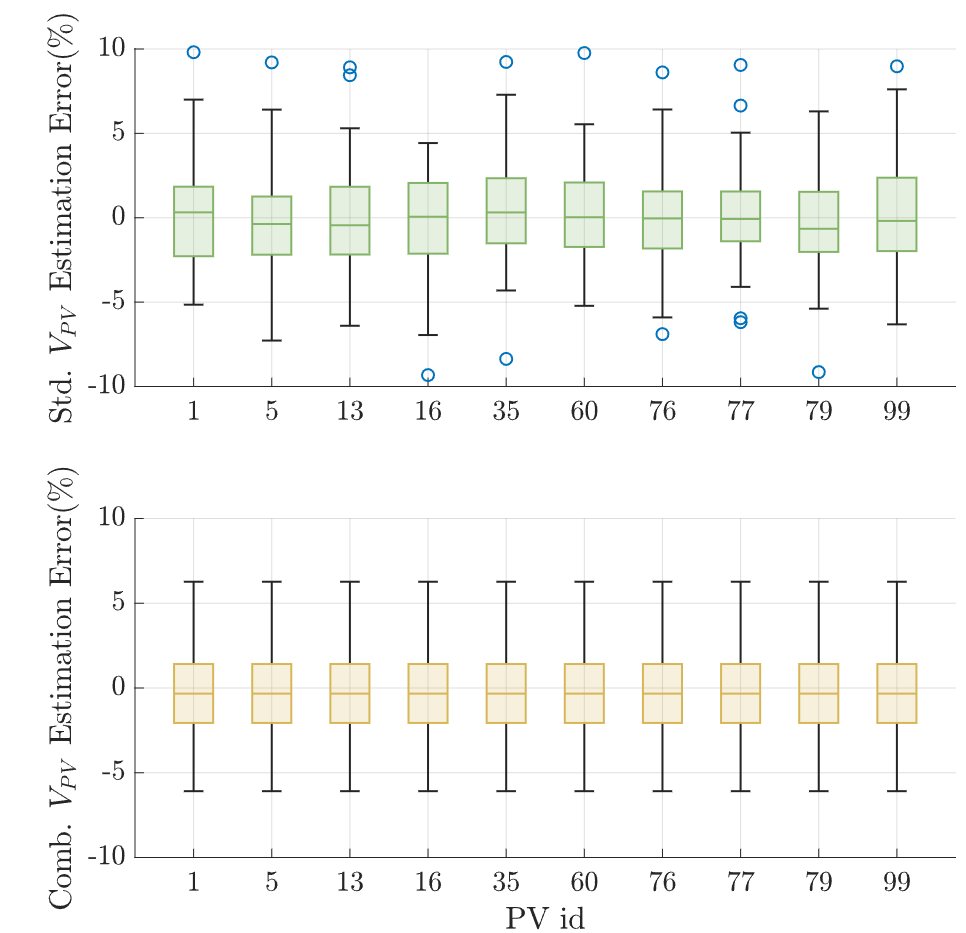}
    \caption{$\VPV{}$ estimation result of 10k node \case{3}{} network without bad data; this box and whisker plot show $\VPV{}$ estimation result of 10 PV systems with the highest error with the \pvse{} and $\cktser{}$. The $\cktser{}$ algorithm (Comb.) estimates are less erroneous and more consistent with no outliers than stand-alone (Std.) estimates.}\label{fig:PV_stdalone_VS_wgrid_SE_10k_top10}
    % \begin{minipage}{9cm}
    % \vspace{0.1cm}
    % \footnotesize 1. This box and whisker plot show $\VPV{}$ estimation result of 10 PV systems with the highest error with the \pvse{} (Std.) and $\cktser{}$ (Comb.) of\case{3}{}\\
    % 2. The $\cktser{}$ algorithm estimates are less erroneous and more consistent with no outliers compared to stand-alone estimates\\
    % \end{minipage}
\end{figure}

\begin{figure}
    \centering
    \includegraphics[scale = 0.35]{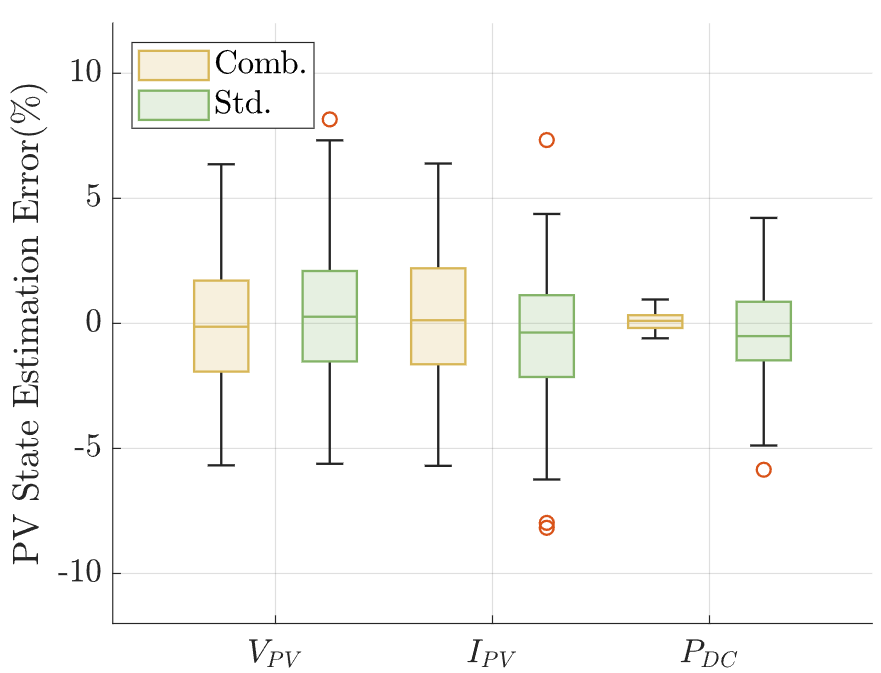}
    \caption{PV state estimation error w/o bad data on \textit{least accurate} PV system in 2869 node \case{2}{a} network. The \pvse{} (Std.) algorithm performs well but is less accurate and has outliers in comparison to the $\cktser{}$ (Comb.) algorithm.}
    \label{fig:AER_case2_ub}
    % \begin{minipage}{9cm}
    % \vspace{0.1cm}
    % \footnotesize The \btse{} (Std.) algorithm NLP performs stable when all parameters are known in the problem, but still has more outliers compared to the $\cktser{}$ (Comb.)
    % \end{minipage}
\end{figure}

Fig. \ref{fig:PV_stdalone_VS_wgrid_SE_10k_top10} plots the DC terminal voltage ($\VPV{}$) estimation error for \textit{least performant} 10 out of the 100 PV systems in the 10k \case{3}{} network. 
The $\cktser{}$ (bottom) observes better estimation accuracy over \pvse{}, which is prone to high-error outliers. 
Fig. \ref{fig:AER_case2_ub} shows the estimation error (\%) for DC terminal voltage ($\VPV{}$), DC current output ($\IPV{}$), and DC power output in 2869 node \case{2}{a}. 
As before, the combined $\cktser{}$ algorithm performs better than the stand-alone \pvse{} algorithm as stand-alone estimation tends to have a wider spread across various noise instances and has outliers. 
We chose this experiment's PV system with the highest power estimation variance.
 
\begin{figure}
    \centering
    \includegraphics[scale = 0.33]{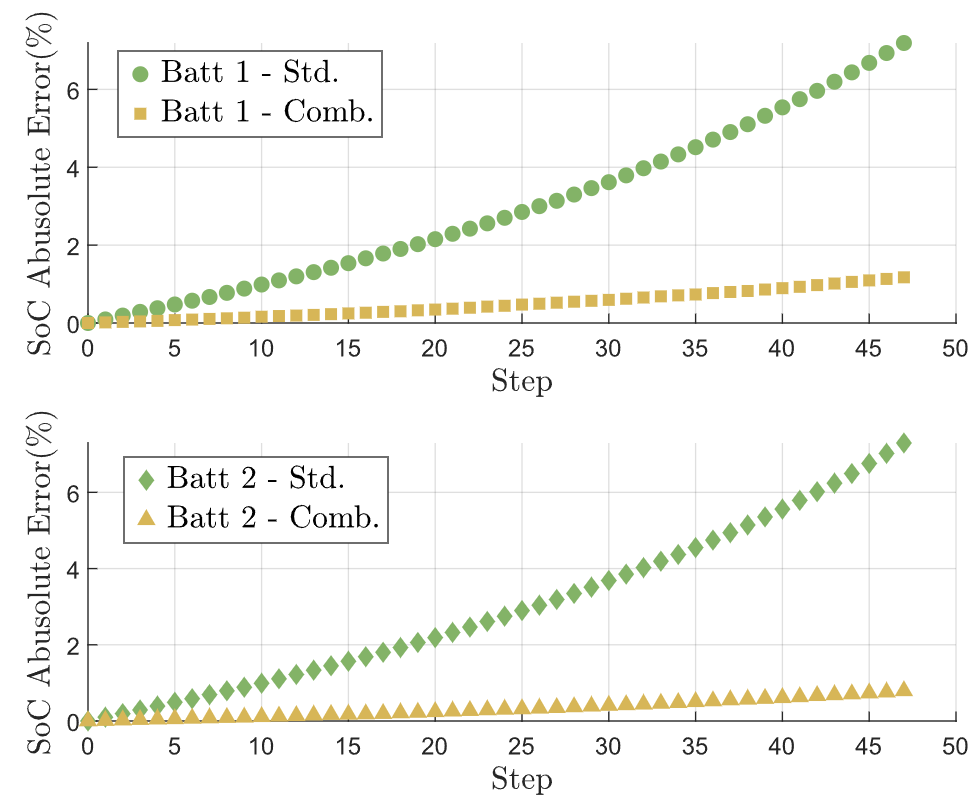}
    \caption{Absolute errors of SoC estimate for two batteries in \case{2}{a} with identical charging and discharging profile and w/o bad-data. 
    Here, each step is 5min, and we simulate a total of 48 steps to represent a 4hr discharging and charging cycle. We observe SoC error accumulates over time and is far worse (\textgreater 6\% vs. \textless 2\%) for the stand-alone \btse{} algorithm.}
    \label{fig:SoC}
    % \begin{minipage}{9cm}
    % \vspace{0.1cm}
    % \footnotesize 1. In this figure, each step is 5min and we simulate a total of 48 steps to represent a 4hr discharging and charging loop of the battery\\
    % \footnotesize 2. The \btse{} algorithm accumulates errors faster compared to $\cktser{}$. Both battery components SoC estimate error breakthrough 6\% for stand-alone algorithm while the combined algorithm keeps below 2\% in this 4hr simulation that both algorithm starts from a TRUE initial SoC
    % \end{minipage}
\end{figure}

For battery systems, in Fig. \ref{fig:SoC}, we compare the state of charge estimation absolute error (\%) for stand-alone \btse{} and combined $\cktser{}$ algorithm for two (2) batteries in \case{2}{a} network. 
Table \ref{tab:State_Est_accuracy_batt} shows estimation results of the DC terminal, open circuit voltages, and DC power states of the batteries in \case{2}{a}. 
While we observe similar results for DC terminal states, SoC errors for stand-alone \btse{} are far higher than combined $\cktser{}$, as they accumulate over time.
% \FloatBarrier
\begin{table}
    \caption{Battery State Estimation Results (w/o bad data)}
    \label{tab:State_Est_accuracy_batt} 
    \centering
    % \hspace*{-0.5cm}
    \scriptsize
    \renewcommand{\arraystretch}{1.2} % Adjust vertical spacing between rows
    \begin{tabular}{@{}l@{\hspace{2pt}}cccccc@{}}
    \hline
    \multirow{2}{*}{\textbf{Algo./Case}} & \multicolumn{2}{c}{$\VBatt{}$} & \multicolumn{2}{c}{$\VOC{}$} & \multicolumn{2}{c}{DC Power} \\ 
    \cmidrule(lr){2-3} \cmidrule(lr){4-5} \cmidrule(lr){6-7}
     &  NRMSE & $\sigma^2_\text{avg}$ &  NRMSE & $\sigma^2_\text{avg}$ & NRMSE & $\sigma^2_\text{avg}$ \\ \hline
    \btse{}/\case{2}{a} & $2.6\text{E-}03$ & $3.3\text{E-}03$ & $3.8\text{E-}02$ & $4.1\text{E-}04$ & $2.4\text{E-}03$ & $7.7\text{E-}07$ \\
    $\cktser{}$/\case{2}{a}     & $2.5\text{E-}03$ & $3.0\text{E-}03$ & $3.8\text{E-}03$ & $4.1\text{E-}04$ & $2.3\text{E-}03$ & $6.9\text{E-}07$ \\ \hline
    \end{tabular}
\end{table}
\subsection{State estimation of solar PV and battery system with bad data}
\begin{figure}
    \centering
    \includegraphics[scale = 0.30]{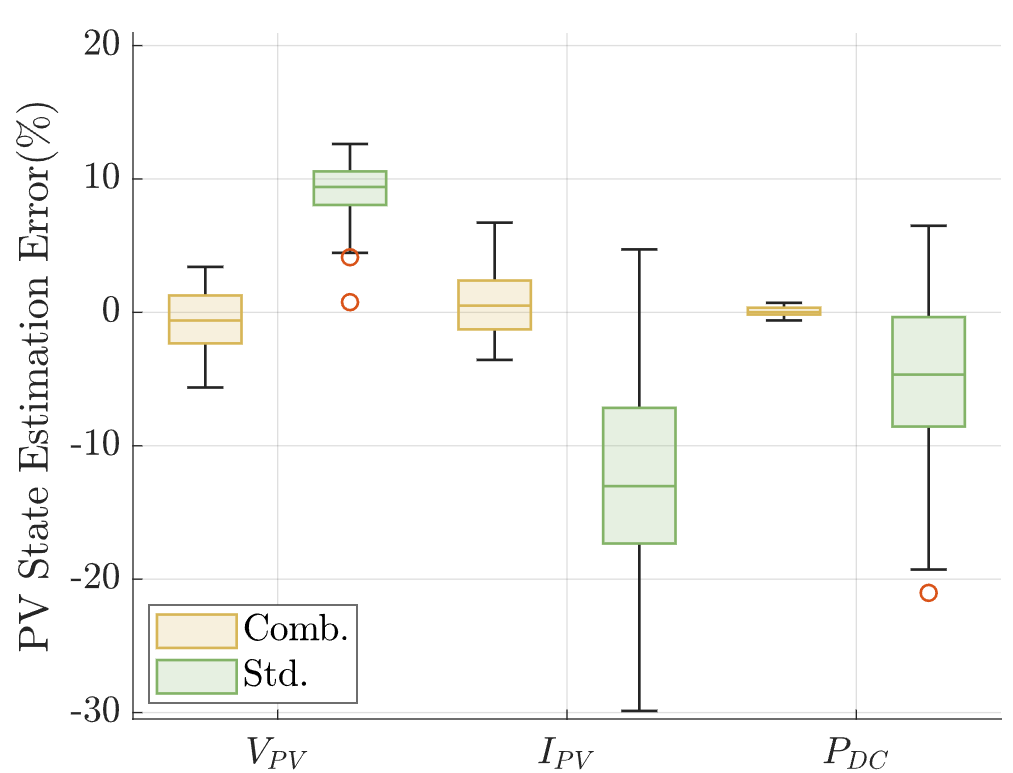}
    \caption{PV state estimation error with bad data in \case{2}{a}. Compared to Fig. \ref{fig:AER_case2_ub}, the estimation quality deteriorates for both algorithms due to biased voltage current readings. However, $\cktser{}$ estimates are significantly more accurate than \pvse{}.}
    \label{fig:AER_case2_bb}
    % \begin{minipage}{9cm}
    % \vspace{0.1cm}
    % \footnotesize 1. Compared to Fig. \ref{fig:AER_case2_ub} due to the biased measurement, the stand-alone estimation deteriorated more by unbiased voltage current readings\\
    % \footnotesize 2. The combined method deteriorates due to bad data for $\VSH{}$ and $\VPV{}$ while affecting less than stand alone method.
    % \end{minipage}
\end{figure}

\begin{figure}
    \centering
    \includegraphics[scale = 0.31]{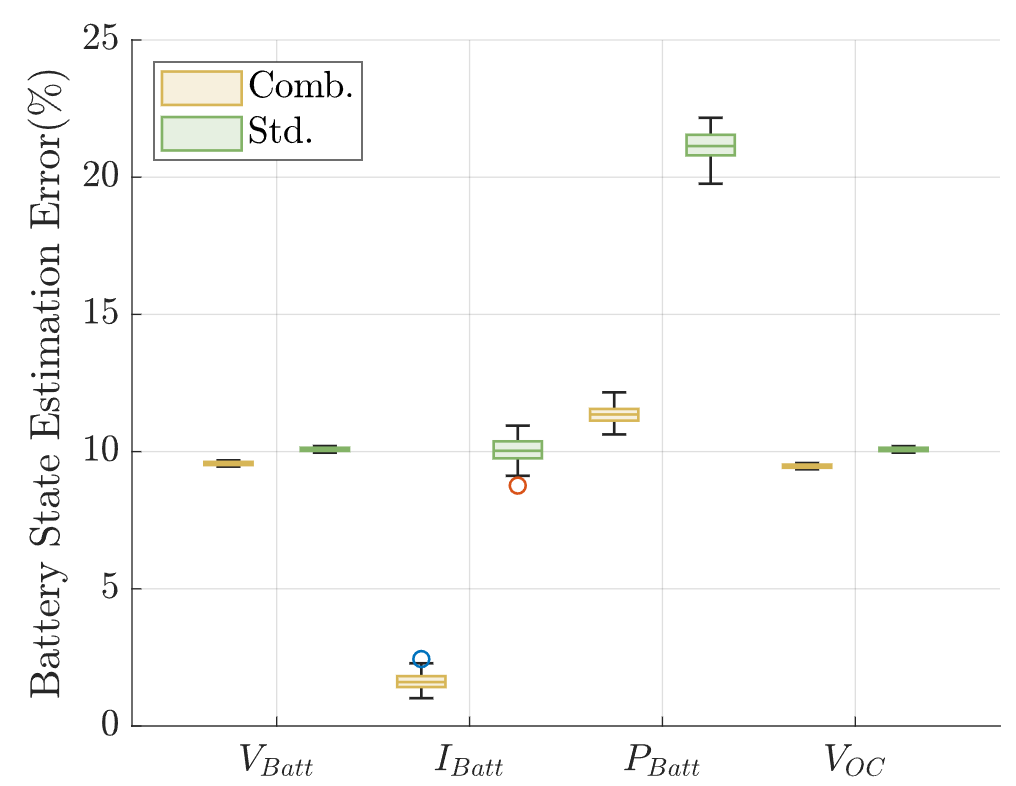}
    \caption{Battery state estimation error with bad data for 2869 node \case{2}{a} network; the x-axis includes various battery states of interest, and the y-axis shows the absolute estimation error. While both algorithms provide biased estimation due to bad data, the $\cktser{}$ method is more accurate compared to \btse{}.}
    \label{fig:AER_case2_bb_batt}
    % \begin{minipage}{9cm}
    % \vspace{0.1cm}
    % \footnotesize 1. The x-axis shows the state of the estimated battery and the y-axis shows the absolute estimation error of a state\\
    % \footnotesize 2. Both algorithms provide biased estimation due to bad data, while the combined method is stronger against bad data 
    % \end{minipage}
\end{figure}

\noindent Realistically, not all measurements are free from bad data. 
Thus, we include bad data points in the next set of Scenario B's experiments. 
Considering the reliability and cost of different measurement equipment, we placed bad data in solar PV and battery measurements instead of the grid measurements as they have cheaper and less reliable meters without a dedicated communication platform. 
We use biased V-I measurements to represent an incorrectly calibrated meter. 
With \case{2}{a} network as the test case, we place a +10\% error on both $\zpvv{}$ and $\zpvi{}$ measurements to create bad data, while the RTU measurements contain only white noise. 
We observe in Fig. \ref{fig:AER_case2_bb}-\ref{fig:AER_case2_bb_batt} and Table \ref{tab:State_Est_accuracy_biased}-\ref{tab:State_Est_accuracy_batt_biased} that for stand-alone (\pvse{} and \btse{}) algorithms, estimates deteriorate significantly.  However, estimates from the $\cktser{}$ algorithm are reasonably accurate as they are aided by redundant and accurate RTU measurements from the grid.

% Table \ref{tab:State_Est_accuracy_biased}-\ref{tab:State_Est_accuracy_batt_biased} summaries PV and battery state estimation results with bad data included.

% \FloatBarrier
\begin{table}[ht]
\caption{PV State Estimation Results (with bad data).}\label{tab:State_Est_accuracy_biased}  
\centering
\scriptsize
\renewcommand{\arraystretch}{1.2} % Adjust the vertical spacing between rows
\begin{tabular}{@{}l@{\hspace{2pt}}cccccc@{}}
\hline
\multirow{2}{*}{\textbf{Algo./Case}} & \multicolumn{2}{c}{$\VPV{}$} & \multicolumn{2}{c}{$\VSH{}$} & \multicolumn{2}{c}{DC Power} \\ 
\cmidrule(lr){2-3} \cmidrule(lr){4-5} \cmidrule(lr){6-7}
 & NRMSE & $\sigma^2_\text{avg}$ & NRMSE & $\sigma^2_\text{avg}$ & NRMSE & $\sigma^2_\text{avg}$ \\ \hline
\btse{}/\case{2}{a} & $8.3\text{E-}02$ & $5.2\text{E-}01$ & $7.3\text{E-}02$ & $3.4\text{E-}01$ & $7.0\text{E-}02$ & $6.0\text{E-}04$ \\
$\cktser{}$/\case{2}{a}    & $2.6\text{E-}02$ & $8.6\text{E-}01$ & $2.3\text{E-}02$ & $7.7\text{E-}01$ & $2.1\text{E-}02$ & $1.2\text{E-}06$ \\ \hline
\end{tabular}
\begin{minipage}{9cm}
\vspace{0.1cm}
\footnotesize Compared to scenario A w/o bad data, both algorithms in scenario B provide less accurate estimates. However, $\cktser{}$ is significantly more accurate, as shown in NRMSE results; the variance of $\VSH{}$ and $\VPV{}$ are higher due to the bad data, especially non-zero mean noise.
\end{minipage}
\end{table}
\begin{table}
    \caption{Battery State Estimation Results (with bad data).}\label{tab:State_Est_accuracy_batt_biased} 
    \centering
    \scriptsize
    \renewcommand{\arraystretch}{1.2} % Adjust vertical spacing between rows
    \begin{tabular}{@{}l@{\hspace{2pt}}cccccc@{}}
    \hline
    \multirow{2}{*}{\textbf{Algo./Case}} & \multicolumn{2}{c}{$\VBatt{}$} & \multicolumn{2}{c}{$\VOC{}$} & \multicolumn{2}{c}{DC Power} \\ 
    \cmidrule(lr){2-3} \cmidrule(lr){4-5} \cmidrule(lr){6-7}
     &  NRMSE & $\sigma^2_\text{avg}$ &  NRMSE & $\sigma^2_\text{avg}$ & NRMSE & $\sigma^2_\text{avg}$ \\ \hline
    \btse{}/\case{2}{a} & $9.2\text{E-}02$ & $9.4\text{E-}03$ & $9.2\text{E-}02$ & $9.7\text{E-}02$ & $1.7\text{E-}01$ & $1.0\text{E-}06$ \\
    $\cktser{}$/\case{2}{a}     & $8.6\text{E-}02$ & $4.4\text{E-}03$ & $8.6\text{E-}03$ & $4.8\text{E-}01$ & $9.9\text{E-}02$ & $7.3\text{E-}07$ \\ \hline
    \end{tabular}
    \begin{minipage}{9cm}
    \vspace{0.1cm}
    \footnotesize Compared to unbiased Scenario A, both algorithms yield less accurate results. However, $\cktser{}$ estimates are 1.5x - 8x better than \btse{} per NRMSE numbers.
    \end{minipage}
\end{table}

\subsection{Parameter State joint estimation}
\begin{figure}
    \centering
    \includegraphics[scale = 0.3]{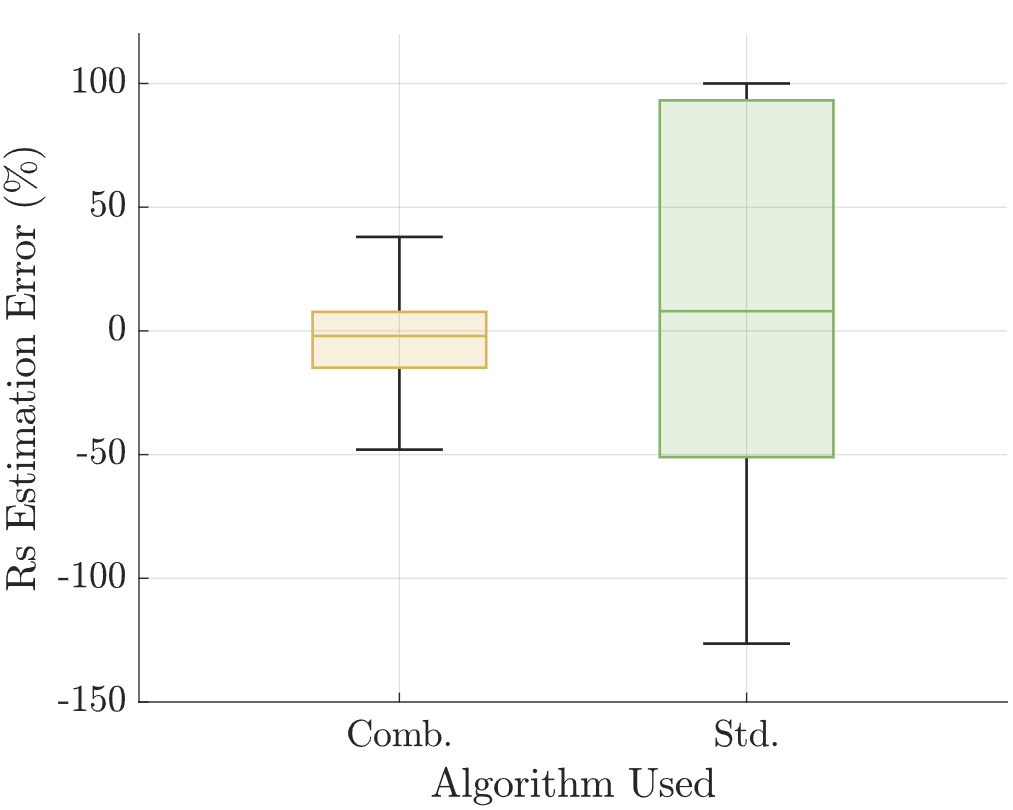}
    \caption{PV parameter estimation result comparison for \case{2}{b}. This box plot shows the parameter estimation results for 100 runs when $\RS{}$ is unknown. $\cktpser{}$ provides an interquartile range between -14\% to 8\% and a median of -2\% compared to the stand-alone method with an interquartile range between -93\% to 51\% and a median of 8\%.}
    \label{fig:Rs_est_compare}
    % \begin{minipage}{9cm}
    % \vspace{0.1cm}
    % \footnotesize 1. This box plot shows the parameter estimation results in 100 runs when $\RS{}$ is unknown. We apply $\cktpser{}$ to this problem\\
    % \footnotesize 2. The combined method provides an interquartile range between -14\% to 8\% and a median of -2\% compared to the stand-alone method with an interquartile range between -93\% to 51\% and a median of 8\% 
    % \end{minipage}
\end{figure}

In scenario C, we consider erroneous parameters in estimation experiments. 
We assume the PV system's $\RS{}$ parameter is unknown for 1 out 10 PV systems and run the stand-alone and combined $\cktpser{}$ algorithm on 2869 node \case{2}{b} network.
Table \ref{tab:Para_Est_accuracy} documents the results. 
%Note that $\cktpser{}$ allows for including erroneous parameters within the optimization framework.
We see that \pvse{} stand-alone algorithm (with unknown parameters) is less accurate compared to $\cktpser{}$ by a factor of $\geq$4x more error.
To visualize the parameter accuracy quality, we build a box plot in Fig. \ref{fig:Rs_est_compare}. 
Our first observation is that $\cktpser{}$ estimation accuracy is significantly better than the stand-alone algorithm.
However, in some scenarios, we still observe up to 40\% error in parameter estimate.
However, note that the median tends to provide a good estimate for the parameter value.
Therefore, if we were to obtain recurrent estimates over a certain time window when the parameter can be assumed unchanged, the median value of the parameter estimate can be expected to be accurate.
% The box plot Fig. \ref{fig:PVPSE_state} shows the estimation accuracy of the PV system states with unknown $\RS{}$ for \pvse{} and $\cktpser{}$ algorithms.
% We repeat the experiment for 100 instances of noise. The plot shows $\cktpser{}$ estimation accuracy is better, especially for $\IPV{}$ and $P_{DC}$, as they are closely coupled to the RTU measurements.
%When multiple data sets of PV systems running in a similar state are available, applying the median estimation of $\RS{}$ can help increase the accuracy as the $\RS{}$ is not volatile.
\begin{table}[ht]
\caption{Parameter Estimation Accuracy Evaluation}\label{tab:Para_Est_accuracy}
    \centering
    \renewcommand{\arraystretch}{1.2} % Adjust the vertical spacing between rows
    \begin{tabular}{@{}lcc@{}}
    \hline
    \textbf{Algo./Case} & $\textbf{NRMSE}_{{\mathcal{P}}}$&$\sigma^2_{{\mathcal{P}}}$\\ 
     \hline
     \pvse{}/\case{2}{b}         &7.1E-01  & 1.2E-01\\
    $\cktpser{}$/\case{2}{b}          & 1.8E-01 & 7.4E-03\\ 
    \hline
\end{tabular}
\end{table}
% \begin{figure}
%     \centering
%     \includegraphics[scale = 0.22]{images/PV_PSE_SE.png}
%     \caption{PV parameter estimation result comparison for \case{2}{b}. The $\cktpser{}$ provides more accurate state estimation in this scenario.}
%     \label{fig:PVPSE_state}
%     % \begin{minipage}{9cm}
%     % \vspace{0.1cm}
%     % \footnotesize 
%     % \end{minipage}
% \end{figure}
% \FloatBarrier
\subsection{Scalability} 
\noindent Next, we study the scalability of combined $\cktser{}$ and $\cktpser{}$ algorithms.
The time-complexity of stand-alone algorithms is less critical to discuss as each component is independent and can be estimated in parallel; the bottleneck will always be the largest subsystem; in our case, the largest subsystem is the grid subsystem.
% \begin{table}
% \caption{Solving time}\label{tab:runtime}
%     \centering
%     \renewcommand{\arraystretch}{1.2} % Adjust the vertical spacing between rows
%     \begin{tabular}{@{}lc@{}}
%     \hline
%     \textbf{Algo./Case} & Time(ms)\\ 
%     \hline
%     $\cktser{}$/\case{1}{}          & 7.2E+01 \\
     
%     $\cktser{}$/\case{2}{a}           &4.0E+03  \\ 
    
%     $\cktpser{}$/\case{2}{b}          & 3.9E+03 \\ 
    
\begin{figure}
    \centering
    \includegraphics[scale = 0.22]{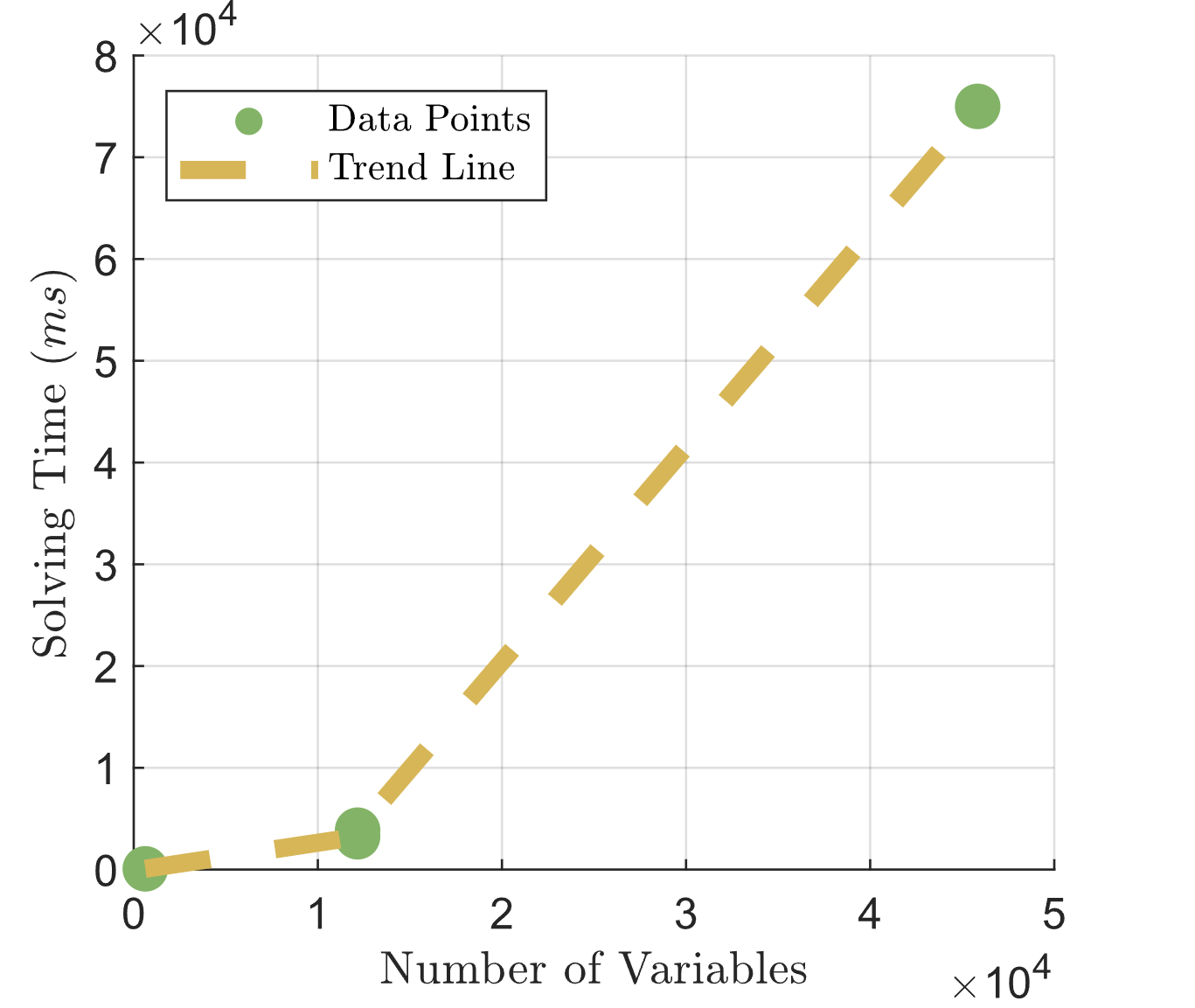}
    \caption{Runtime for different cases v.s. number of variables in the case. The number on the x-axis shows the number of unknown variables in the optimization problem.}
    \label{fig:Runtime}
    \begin{minipage}{9cm}
    \vspace{0.1cm}
    \footnotesize
    \end{minipage}
\end{figure}

%     $\cktser{}$/\case{3}{}          & 7.5E+05 \\ 
%     \hline
% \end{tabular}
% \begin{minipage}{9cm}
% \vspace{0.1cm}
% \footnotesize  $\cktser{}$/\case{2}{a} and $\cktpser{}$/\case{2}{b} are different in solving time due to the unknown parameter in the $\cktpser{}$/\case{2}{b}.
% \end{minipage}
% \end{table}
The time complexity of the combined $\cktser{}$ algorithm is shown in Fig. \ref{fig:Runtime}.
The smaller networks with up to 10 PV and battery systems solve in 4 seconds.
The largest $>$10k nodes \case{3}{} network with 40k+ variables and 100 PV systems takes 80 seconds at most.
This validates that our approach is scalable and can be implemented in grid control rooms.
%the computation speed depends more on the speed of solving the optimization problem, the plot is as shown in Fig.  We can solve a 10k bus system with more than 40k variables within 80 seconds.

\section{Conclusion}
% \begin{itemize}
    % \item [1] The circuit method can incorporate both AC and DC in one formulation, allowing us to include extra measurements from solar PV and battery storage system
    % \item [2] The circuit method can be subtracted or increased by substitution theorem, more measurements can be introduced whenever available in this formulation with the cost of more complicated circuit and longer computation time.
    % % \TODO{It is important to show this method can be expanded with more measurements and more subcircuit, cuz the paper covered the non-exactness of simply using PV or PQ to represent the HES system, then representing the whole system with one equivalent circuit also seems to be quit general.}
    % \item [3] Compare the separate estimation and co-estimation
% \end{itemize}
% \noindent To sum up, the conducted experiment shows that the $\cktser{}$ and $\cktpser{}$ are scaleable algorithms that can incorporate additional measurements and estimate the states of interest up to 10k bus systems and more than 40k variables. $\cktpser{}$ is more robust against unknown parameters compared to \pvse{} and \btse{}.
\noindent The paper was inspired by the lack of standard techniques to estimate utility-scale solar PV and battery storage system states consistent with transmission system measurements amid the growing number of these systems on the grid. 
We built a circuit-based joint parameter-state estimation framework $\cktpser{}$ to accurately model and estimate the states for solar PV, battery storage, and the transmission grid systems within one combined framework.
We leveraged circuit models for solar PV, battery storage, and grid components and combined them into one aggregated circuit to perform estimation analysis.
We included measurements from all components without loss of generality, and we minimized the measurement noise subject to circuit-based Kirchhoff's constraints to obtain state estimates. We draw the following conclusions:
\begin{itemize}
    \item Combined $\cktser{}$ algorithm outperforms stand-alone algorithms regarding estimation accuracy with or without the existence of bad data; we observe \textgreater 3x improvement in scenario B for PV systems and 1.5x - 8x improvement for battery per standard error metrics
    %It can provide a more accurate estimation, especially in scenarios the renewable energy systems have poor measurement accuracy
    \item Combined $\cktpser{}$ is more robust to erroneous parameters. $\cktpser{}$ is 3.9x more accurate per standard error metric compared to stand-alone algorithms.
    %Stand-alone algorithms are capable of performing a good accuracy when measurements are good, thus we suggest using \pvse{}, \btse{} for its simplicity when good measurements are available and keep $\cktser{}$ as a reference as it is more expensive to compute
    \item  Combined $\cktpser{}$ algorithm scales and can be used in grid control rooms; we solve the estimation problem for $>$10k node network with 100 PV systems in about 80 seconds
    %performs a more reliable parameter estimation with an accurate median, yet has up to 40\% error in some scenarios. Therefore we suggest recurrent estimates over a time period with an unchanged parameter and use the median value of the estimates.
\end{itemize}
%As a final note, our experiment showcases that $\cktpser{}$ can estimate a realistic-sized network within 80 seconds, making it applicable in the control room. 

\newpage
\printbibliography
\end{document}